\documentclass{aa}
\usepackage[colorlinks=true,     linkcolor=blue, citecolor=blue, filecolor=blue, urlcolor=blue]{hyperref}
\usepackage{graphicx}
\usepackage{txfonts}
\usepackage{amsmath}
\usepackage[section]{placeins}
\FloatBarrier
\usepackage[export]{adjustbox}
\usepackage[version=4]{mhchem}
\usepackage[dvipsnames]{xcolor}
\begin{document}

\newcommand{\dal}[1]{\textcolor{purple}{#1$^{(DAL)}$}}
\newcommand{\Teff}{$T_{\rm{eff}}$}
\newcommand{\PT}{$P_{\mathrm{gas}}-T_{\mathrm{gas}}$}

\title{The \texttt{MSG} model for cloudy substellar atmospheres:}
\subtitle{A grid of self-consistent substellar atmosphere models with microphysical cloud formation}
\titlerunning{\texttt{MSG} cloudy substellar atmospheres}
\authorrunning{Campos Estrada et al.}

\author{Beatriz Campos Estrada
          \inst{1,2,3,4}\fnmsep\thanks{E-mail: becampos@mpia.de}, David A. Lewis\inst{2,3,5,6,7}, Christiane Helling\inst{2,3}, Richard A. Booth\inst{8}, \\ Francisco Ard\'evol Mart\'inez\inst{9,10,11,7}, Uffe G. J{\o}rgensen\inst{1}
          }

\institute{Centre for ExoLife Sciences, Niels Bohr Institute,  {\O}ster Voldgade 5, 1350 Copenhagen, Denmark
         \and
             Space Research Institute, Austrian Academy of Sciences, Schmiedlstrasse 6, A-8042 Graz, Austria
             \and
             TU Graz, Fakultät für Mathematik, Physik und Geodäsie, Petersgasse 16, A-8010 Graz, Austria 
              \and 
              Max-Planck-Institut für Astronomie, Königstuhl 17, D-69117 Heidelberg, Germany
              \and
              Centre for Science at Extreme Conditions, School of Physics and Astronomy, The University of Edinburgh, Edinburgh, UK
              \and
              Scottish Universities Physics Alliance, School of Physics and Astronomy, The University of Edinburgh, Edinburgh, UK
              \and 
              School of GeoSciences, The University of Edinburgh, Edinburgh, UK
              \and
             School of Physics and Astronomy, University of Leeds, Leeds LS2 9JT, UK
             \and
             Kapteyn Astronomical Institute, University of Groningen, Groningen, The Netherlands
            \and
             Netherlands Space Research Institute (SRON), Leiden, The Netherlands
            \and
            Centre for Exoplanet Science, The University of Edinburgh, Edinburgh, UK
            }

\date{Received 19 July 2024; Accepted 8 January 2025}

\abstract
    {State-of-the-art JWST observations are unveiling unprecedented views into the atmospheres of substellar objects in the infrared, further highlighting the importance of clouds. Current forward models struggle to fit the silicate clouds absorption feature at $\sim10\,\mu$m observed in substellar atmospheres.}
    {In the \texttt{MSG} model, we aim to couple the \texttt{MARCS} 1D radiative-convective equilibrium atmosphere model with the 1D kinetic, stationary, non-equilibrium, cloud formation model \texttt{DRIFT}, also known as \texttt{StaticWeather}, to create a new grid of self-consistent cloudy substellar atmosphere models with microphysical cloud formation. We aim to test if this new grid is able to reproduce the silicate cloud absorption feature at $\sim10\,\mu$m.}
    {We model substellar atmospheres with effective temperatures in the range $T_{\rm{eff}}$ = 1200-2500 K and with $\log(g)=4.0$. We compute atmospheric structures that self-consistently account for condensate cloud opacities based on microphysical properties. We present an algorithm based on control theory to help converge such self-consistent models. Synthetic atmosphere spectra are computed for each model to explore the observable impact of the cloud microphysics. We additionally explore the impact of choosing different nucleation species (\ce{TiO2} or \ce{SiO}) and the effect of less efficient atmospheric mixing on these spectra.}
    {The new \texttt{MSG} cloudy grid using \ce{TiO2} nucleation shows spectra which are redder in the near-infrared compared to the currently known population of substellar atmospheres. We find the models with \ce{SiO} nucleation, and models with reduced mixing efficiency are less red in the near-infrared.}
    {We present a new grid of \texttt{MSG} models for cloudy substellar atmospheres that include cloud-radiative feedback from microphysical clouds. The grid is unable to reproduce the silicate features similar to those found in recent JWST observations and \textit{Spitzer} archival data. We thoroughly discuss further work that may better approximate the impact of convection in cloud-forming regions and steps that may help resolve the silicate cloud feature.}

%\keywords{  }

   \maketitle
%
%________________________________________________________________

\section{Introduction}

Mineral clouds in brown dwarf atmospheres were first proposed almost 40 years ago by \citet{Lunine1986}. By comparing modelled temperature-pressure profiles with the condensation curves of some refractory materials, several studies reached the conclusion that mineral clouds, composed of silicates, metals and aluminium oxides, should exist in substellar atmospheres \citep[e.g.][]{Lunine1986, Burrows97, Chabrier2000, Marley2002}. 

During these past 40 years, several different models have been developed and used to model these cloudy atmospheres. Cloud models used in the present day tend to either be based on a parameterised approach or in a more complex approach which treats cloud formation microphysical processes kinetically. The parameterised approaches generally handle cloud particle compositions calculated from thermochemical equilibrium and either assume an average cloud particle size or define a specific size distribution for the cloud particles. This allows for a determination of the cloud optical properties based on Mie theory \citep{Mie1908}. The differences between parameterised models arise from the different manner in which they parameterise microphysical processes. For example, the \citet{Ackerman2001} approach is based on determining cloud distributions by balancing particle sedimentation with vertical mixing, with the vertical extent of the clouds being controlled by a sedimentation efficiency parameter. On the other hand, the \citet{Cooper2003} approach obtains the average cloud particle size by balancing the timescales of microphysical processes following \citet{Rossow1978}. The microphysical approach considers complex processes that lead to cloud formation, such as nucleation, condensation, evaporation, and transport. Within the microphysical approach, there are two major modelling methods, the bin-method \citep[e.g.][]{Toon1979, Ohno2017, Gao2018, Powell2018, Kawashima2018} and the dust moments method \citep[e.g.][]{Gail1988, Dominik1993, Woitke2003, Woitke2004, Helling2006}{}{}. For a comparison and a review of the different cloud modelling methodologies and their advantages and disadvantages see \citet{Helling2008_comparison} and \citet{Gao2021}.

To consider the effect of the cloud's radiative feedback onto substellar atmospheres, we must employ self-consistent models which take into consideration the effects of the cloud opacity and gas-phase element depletion caused by the cloud formation onto the atmospheric structure. Using brown dwarfs to assess the implications of cloud formation in substellar atmospheres has a major advantage over using irradiated planets: one does not need to model the external radiation field of a host-star, which immensely simplifies the radiative-transfer problem, and there is no brightness from a host hindering the observations. Modelling and observing brown dwarf atmospheres will still increase our understanding of extrasolar giant planet (EGP) atmospheres. This is because L- and T- dwarfs are analogues of EGPs as they share temperatures, surface gravities and ages. 

\citet{Cushing2006} reported what can be seen as the very first evidence for silicate cloud features from \textit{Spitzer} mid-infrared observations of several L-dwarfs. More recently, \citet{Suarez2022} analysed hundreds of M- to T- dwarfs \textit{Spitzer} infrared spectra. They found the silicate cloud absorption feature to be fairly common in L-dwarfs. However, not all of the L-dwarf spectra showed silicate absorption features. In addition to this, performing retrievals on archival data, \citet{Burningham2021} and \citet{Vos2023} find evidence for silicate clouds in three different brown dwarfs.

We have now entered the JWST era which has provided us with a new view of substellar atmospheres. The JWST MIRI instrument allows for medium resolution (R$\sim$1500 - 3250) spectroscopy measurements in the mid-infrared (MIR). A silicate cloud absorption feature was detected with JWST MIRI in the planetary mass companion VHS\,1256\,b, between 9 and 11\,$\mu$m by \citet{Miles2023}. In addition to this, we have had the first confirmation of the presence of quartz clouds in a hot Jupiter, WASP-17\,b \citep{Grant2023}, and in a warm Neptune, WASP-107\,b \citep{Dyrek2024}.

\citet{Petrus2024} have presented an analysis of the VHS\,1256\,b data using 5 different forward-models: \texttt{ATMO} \citep{Tremblin2015}, \texttt{Exo-REM} \citep{Charnay2018}, \texttt{Sonora Diamondback} \citep{Morley2024}, \texttt{BT-Settl} \citep{Allard2012} and \texttt{DRIFT-PHOENIX} \citep{Helling2008_consistent, Witte2009, Witte2011}. 
Out of the five models, \texttt{ATMO} is the only that is cloudless. \texttt{BT-Settl} and \texttt{DRIFT-PHOENIX} consider cloud microphysics self-consistently, although based on different modelling approaches. \texttt{Exo-REM} makes a parameterisation of the cloud microphysics processes, in a self-consistent manner. \texttt{Sonora Diamondback} is self-consistent; however, it does not use cloud microphysics and instead uses the \citet{Ackerman2001} approach to consider the presence of clouds in the atmosphere. None of the five models can reproduce the silicate cloud absorption feature of VHS\,1256\,b \citep{Petrus2024}. 

There is a need for an updated self-consistent grid with cloud microphysics. Since the publication of the \texttt{DRIFT-PHOENIX} grid over 10 years ago, a considerable number of important molecules and atoms, such as \ce{CH4} \citep{CH4_Yurchenko2017}, \ce{NH3} \citep{NH3_Coles}, \ce{TiO} \citep{TiO_McKemmish}, \ce{VO} \citep{VO_McKemmish}, Na \citep{Allard2019}, K \citep{Allard2019}, have had their line-lists updated. In addition to this, equilibrium chemistry models have become more complex but also computationally faster. In this new \texttt{MSG} grid, we include five more cloud species (\ce{SiO}[s], \ce{CaTiO3}[s], FeO[s], FeS[s] and \ce{Fe2O3}[s]) than those considered in \texttt{DRIFT-PHOENIX} (\ce{TiO2}[s], \ce{Al2O3}[s], Fe[s], \ce{SiO2}[s], MgO[s], \ce{MgSiO3}[s] and \ce{Mg2SiO4}[s]). The formulation used to treat atmospheric mixing has also been updated and is explained in detail in section~\ref{sec:SW}.
\texttt{MSG} grid combines the \texttt{MARCS} atmosphere model, with the equilibrium chemistry model \texttt{GGchem} and the \texttt{DRIFT} cloud formation model. Here, we have coupled \texttt{MARCS} to \texttt{DRIFT} using a new algorithm which ensures convergence of the cloud opacity and gas-phase element depletion caused by the cloud formation along the atmosphere.

%__________________________________________________________________
\texttt{MARCS}  was originally written to model the atmospheres of solar-type stars \citep[][]{Gustafsson1975}{}{} and has since been extended to model stellar atmospheres ranging from late A-type to early M-type stars \citep[e.g.][]{Lambert1986, Jorgensen1992, Gustafsson2008}{}{}. \texttt{MARCS} has been used for multiple purposes from abundance analyses \citep[e.g.][]{Blackwell1995, SiqueiraMello2016}{}{}, to $\rm{H_2O}$ detections \citep[][]{Ryde2002,  Aringer2002}{}{}, to modelling carbon stars and white dwarf atmospheres \citep{Jorgensen1992, Jorgensen2000} and instrument calibrations \citep[][]{Decin2003, Decin2007}{}{}. \texttt{MARCS} stellar atmosphere models are being used to compute stellar parameters of PLATO targets \citep{Gent2022, Morello2022}. A summary of the development of \texttt{MARCS} can be found in \citet{Gustafsson2008}. 

More recently, \texttt{MARCS} has been expanded to model the cloudy atmospheres of late-type M-dwarfs and early-type L-dwarfs \citep[][]{Juncher2017}{}{}. However, this extension was limited to effective temperatures down to 2000\,K due to convergence complications. The convergence issues primarily arose from \texttt{MARCS} using the electron pressure as an independent variable (instead of the gas pressure as it is commonly used in other models). Addressing this challenge, \citet{MSGpaper} have adapted the \texttt{MARCS} framework to account for the extremely low abundance of free electrons at cooler temperatures, successfully resolving these convergence issues.
In this work, following the new \texttt{MSG} grid (based on \texttt{MARCS}) presented in \citet{MSGpaper}, we expand the \texttt{MSG} grid of cloudy substellar objects down to effective temperatures of 1200\,K.

\texttt{DRIFT}, also known as \texttt{StaticWeather}, is a 1D non-equilibrium, stationary, microphysical model which kinetically treats cloud formation \citep{Woitke2003,Woitke2004,Helling2006,Helling2008}. \texttt{DRIFT} models several key cloud formation processes, including nucleation, bulk growth, evaporation, gravitational settling of cloud particles, and the depletion of gas-phase element abundances. The model further employs a parameterised scheme for atmospheric mixing that acts to replenish depleted element abundances and counteract the gravitational settling of the cloud particles. The gas-phase composition is computed with the equilibrium chemistry code \texttt{GGchem} \citep{Woitke2018}. \texttt{DRIFT} has been applied across a broad range of substellar atmospheres \citep{Helling2006,Helling2008_comparison} and has previously been coupled to the \texttt{PHOENIX} code \citep{Hauschildt1999} to produce the \texttt{DRIFT-PHOENIX} atmosphere model grid and synthetic spectra \citep{Helling2008_consistent,Witte2009,Witte2011}. Recent works have explored cloud formation in exoplanet atmospheres utilising a hierarchical modelling approach of post-processing \texttt{DRIFT} onto 3D cloud-free General Circulation Models (GCMs). This has been applied to hot and ultra-hot Jupiter atmospheres such as WASP-18 b \citep{Helling2019_WASP18b}, WASP-43 b \citep{Helling2020_WASP43b,Helling2021_WASP_Comparison}, HAT-P-7 b \citep{Helling2019_HATP7b,Helling2021_WASP_Comparison}, and WASP-96 b \citep{Samra2023_WASP96b}, as well as to a grid of model exoplanet atmospheres spanning a wide physical parameter space \citep{Helling2023_gridpaper}.

We start by describing cloud formation and the \texttt{DRIFT} model in section~\ref{sec:SW}. Next, we describe atmosphere modelling with \texttt{MARCS} in section~\ref{sec:marcs}. In section~\ref{sec:sw_marcs} we describe how we couple \texttt{MARCS} to \texttt{DRIFT}, in order to compute the new \texttt{MSG} model grid. In section~\ref{sec:results} we present our results. We start by presenting an overview of the grid. This includes an overview of the pressure-temperature profiles, the cloud properties and the model spectra. In section~\ref{sec:nucleation} we explore the effect of changing the cloud condensation nuclei (CCN) from \ce{TiO2} to \ce{SiO}. We investigate the effect of scaling down the mixing efficiency in section~\ref{sec:mixing}. Section~\ref{sec:discussion} discusses our results, including our prospects for future cloudy models using the \texttt{MSG} grid. 

\section{Cloud formation and the \texttt{DRIFT} model}
\label{sec:SW}
The \texttt{DRIFT} model addresses cloud formation by considering the formation of cloud condensation nuclei (CCN), i.e. nucleation, followed by their growth and evaporation. This involves solving a system of dust moment and element conservation equations. Additionally, the model accounts for gravitational settling and element replenishment through convective overshooting. 

Cloud formation starts with the emergence of CCN, on which all thermally stable cloud species may grow through gas-surface reactions \citep[e.g.][]{Helling2019}. In \texttt{DRIFT}, the nucleation rate of CCN is considered by applying the modified classical nucleation theory of \citet{Gail1984}, as described in \citet{Woitke2004}. Dust growth of mixed-material cloud particles is computed following \citet{Woitke2003}, \citet{Helling2006} and \citet{Helling2008}. The cloud particles will gravitationally settle and, as the cloud particles fall, they will encounter different temperatures and gas densities, which causes their composition to change due to the changing thermal stability of the cloud species. The cloud particles continue to grow as they fall until the temperature is so high that all considered cloud species begin to evaporate. The growth and evaporation processes change the local gas-phase element abundances because elements which participate in the cloud formation are depleted. An element replenishment mechanism must exist for a cloud layer to persist (i.e., a source of fresh, non-depleted elements must exist). Here, the replenishment of elements is considered via convective overshooting, which is described in more detail later in this section.

In this work, we model plane-parallel, quasi-static substellar atmospheres. Following this, the evolution of the cloud particles can be described by a series of moment equations for mixed-material cloud particles \citep{Gail1988, Dominik1993, Woitke2003, Helling2006}, these are
\begin{equation}
  - \frac{d}{dz} \left(\frac{\rho_d}{c_{\rm{T}}}L_{j+1} \right) = \frac{1}{\xi_{\rm{1Kn}}} \left( - \frac{\rho L_j}{\tau_{\rm{mix}}} + V_\ell^{j/3}\,J_\star\,+ \frac{j}{3} \, \chi_{\rm{net}} \, \rho L_{j-1}\right)
  \label{eq:moment}
\end{equation}
for $j$=(0, 1, 2), where $V_\ell$ is the minimum volume of a cluster to be considered a cloud particle, $\rho$ is the gas mass density, $\rho_d$ is the cloud mass density, $J_\star [\rm{cm^{-3}s^{-1}}]$ is the total nucleation rate, $\chi_{\rm{net}} [\rm{cm \, s^{-1}}]$ is the net growth velocity, $\xi_{\rm{1Kn}} [\rm{dyn \, cm^{-3}}]$ is the drag force density, $c_{\rm{T}}$ is a mean thermal velocity and $\tau_{\rm{mix}}$ is a mixing timescale. For more details on Equation~\ref{eq:moment} see \citet{Woitke2003,Woitke2004} and \citet{Helling2008}. The moments $L_j\,[\rm{cm}^j/\rm{g}] $ of the cloud particle volume distribution function $f(V)\,[\rm{cm}^{-6}]$ are defined as
\begin{equation}
\rho\,L_j = \int_{V_\ell}^{\infty} f(V)\,V^{j/3}\,dV.
\label{eq:dust_mom}
\end{equation}

The total cloud particle volume per $\rm{cm^3}$ of total matter is determined by the 3$^{\rm{rd}}$ dust moment, $L_3$, as
\begin{equation}
    \rho\,L_3 = \int_{V_\ell}^{\infty}\, f(V)\,V\,dV = V_{\rm{tot}}\,\,[\rm{cm}^3 \rm{cm}^{-3}] 
\end{equation}
where $V$ is the volume of an individual dust particle, and $V_\ell$ is the lower integration boundary. Similarly, we can define the volume $V_s$ of a certain cloud species $s$ by
\begin{equation}
\label{eq:l3s}
    \rho\,L_3^s = \int_{V_\ell}^{\infty}\, f(V)\,V^s\,dV = V_s \,\, [\rm{cm}^3 \rm{cm}^{-3}]
\end{equation}
where $V^s$  is the sum of island volumes of cloud species $s$ in an individual dust particle, $V_{\mathrm{tot}} = \sum V_s$, and $L_3 =\sum L_3^s$. 
Because we consider multiple cloud species, we must use a set of equations for mixed-material cloud particles, i.e. the third dust moment equations for all volume contributions (one for each cloud species $s$). Following equations~\ref{eq:moment} and~\ref{eq:l3s}, one finds
\begin{equation}
\label{eq:l4s}
  - \frac{d}{dz} \left(\frac{\rho_d}{c_{\rm{T}}}L_{4}^{s} \right) = \frac{1}{\xi_{\rm{1Kn}}} \left( - \frac{\rho L_3^s}{\tau_{\rm{mix}}} + V_\ell^{s}\,J_\star\,+ \frac{j}{3} \, \chi_{\rm{net}}^s \rho L_{2}\right),  
\end{equation}
where $L_4^s$ is defined as $L_4^s = L_4\,V_s/V_{\mathrm{tot}}$. For the full derivation of Equation~\ref{eq:l4s} see \citet{Helling2008}.

Equations~\ref{eq:moment} for $j \in (0, 1, 2)$ and equations~\ref{eq:l4s} for $s \in (0,1,2,...,S)$, 
where $S$ is the number of cloud species considered, form a system of $(S+3)$ ordinary differential equations (ODEs) for the unknowns $(L_1, L_2, L_3^s, L_4^s)$. 

The gas-phase element abundance conservation as affected by nucleation, growth and evaporation is given by \citep{Woitke2004, Helling2008},
\begin{multline}
    \frac{n_{\langle H \rangle}\,(\epsilon^0_i - \epsilon_i)}{\tau_{\rm{mix}}} = \nu_{i,0}\,N_\ell\,J_\ast \,  \\
     + \, \sqrt[3]{36\pi} \,\rho_g\,L_2\,  \sum_{r=1}^{R} \frac{\nu_{i,s}n^{\rm{key}}_r \upsilon^{\rm{rel}}_r \alpha_r}{\nu^{\rm{key}}_r} \left(1-\frac{1}{S_r\,b^{s}_{\rm{surf}}}\right).
\label{eq:elem_cons}
\end{multline}
The term on the l.h.s. describes the gas-phase element replenishment through atmospheric mixing, where $n_{\langle H \rangle}$ is the total hydrogen nuclei density, and $\epsilon^0_i$ and $\epsilon_i$ are the initial and depleted element abundances of element $i$ normalised to hydrogen, respectively. The first term on the r.h.s. describes the element depletion by nucleation, where $\nu_{i,0}$ is the stoichiometric coefficient of the CCN and $N_\ell$ is the number of monomers in the CCN. The second term on the r.h.s. describes the element depletion by evaporation/growth, where $\rho$ is the gas mass density, $L_2$ is the 2$^{\rm{nd}}$ dust moment as defined in Equation~\ref{eq:dust_mom}, $r$ is the index for the chemical surface reaction, $\nu_{i,s}$ is the stoichiometric coefficient of element $i$ in solid material $s$, $n^{\rm{key}}_r$ is the particle number density of the key reactant in the gas-phase, $\upsilon^{\rm{rel}}_r$ is the relative thermal velocity of the gas species taking part in reaction $r$, $\alpha_r$ is the sticking coefficient of reaction $r$, $\nu^{\rm{key}}_r$ is the stoichiometric factor of the key reactant in reaction $r$, $S_r$ is the reaction's supersaturation ratio and $1/ b^{s}_{\rm{surf}} = V_s / V_{\rm{tot}}$ is the volume ratio of solid $s$ to the total cloud particle volume $V_{\rm{tot}}$. For more details on Equation~\ref{eq:elem_cons} see \citet{Woitke2004} and \citet{Helling2008}. In this work $N_l = 1000$ and $\alpha_r =1$ for all reactions.

The element conservation equations~\ref{eq:elem_cons} provide algebraic auxiliary conditions for the ODE system (Equations~\ref{eq:moment} and~\ref{eq:l4s}). First, the system of non-linear algebraic equations~\ref{eq:elem_cons} has to be solved for $\epsilon_i$ at given ($L_2, L_4^s$) before the r.h.s of the ODEs can be computed. The dust volume composition $b^{s}_{\mathrm{surf}}$ is obtained from $L_4^s = L_4/ b^{s}_{\mathrm{surf}}$. The abundance $\epsilon_i$ is strongly dependent on $J_{\star}$, $n^{\rm{key}}_r$ and $S_r$ and therefore an intricate iterative procedure is required to solve the equations. $L_0$ is set by a closure condition \citep[see Appendix A in][]{Helling2008}, $L_{1,2,3}$ are determined by solving for $L_j$ in Equation~\ref{eq:dust_mom} (see \citet{Woitke2003} for details), and $L_4$ is solved by using $L_3$. The numerical methods and iterative processes used in \texttt{DRIFT} are described in detail in 
\citet{Woitke2004}.

The cloud particle number density, $n_d \,[\rm{cm^{-3}}]$, and the average cloud particle size, $\langle a \rangle \, [\rm{cm}]$, can be calculated from the moments \citep{Helling2013} by
\begin{eqnarray}
n_d &=& \rho\,L_0, \\
\langle a \rangle &=& \left( \frac{3}{4\pi}\right)^{1/3}\frac{L_1}{L_0},
\end{eqnarray}
respectively.

Atmospheric mixing is parameterised within the model using a mixing timescale, $\tau_{\rm mix}$. This timescale is introduced by \citet{Woitke2004} as the atmosphere would remain cloud-free in the truly static case. This approach is simplified and assumes the gas/cloud particles mix at a height $z$ is exchanged by cloud-free gas from the deep atmosphere with element abundances $\epsilon_i^0$ on a mixing
timescale $\tau_{\mathrm{mix}}(z)$ (overshooting). Similar to previous works  \cite[e.g.][]{Helling2008, Witte2009} we parameterise $\tau_{\rm mix}$ following the convective mixing and overshooting assumption. In the deepest, convective, atmospheric layers (i.e. the bottom of the atmosphere) we compute $\tau_{mix}$, as
\begin{equation}
    \tau_{\rm{mix}} = \frac{\alpha \, H_p}{v_{\rm{conv}}},
\label{eq:tau_mix}
\end{equation}
where $\alpha$ and $v_{\rm{conv}}$ are the mixing length parameter and the convective velocity respectively, as defined in \citet{Gustafsson2008}. $H_p$ is the atmospheric layer's scale height given by
\begin{equation}
    H_p = \frac{k_B \, T}{m_u \, u \, g},
\end{equation}
where $k_B$ is the Boltzmann's constant, $T$ is the atmospheric layer's temperature, $m_u$ is the mean molecular mass in atomic mass units, $u$ is an atomic mass unit and $g$ is the acceleration of gravity. If a detached convective layer exists higher in the atmosphere, $\tau_{\rm mix}$ is set to constant to the value at the top of the radiative zone just below. We discuss the validity of this assumption in section~\ref{sec:conv}.

In the radiative zone(s), the convective velocity is zero, and therefore, we must parameterise the mixing timescale differently. At each layer, we calculate $\beta$, defined as
\begin{equation}
    \beta = \frac{\rm{log}\,\tau_{\rm{mix}}^{i+1} - \rm{log}\,\tau_{\rm{mix}}^{i}}{\rm{log} \, P_i - \rm{log} \, P_{i+1}}.
\end{equation}
When the calculated $\beta$ exceeds a critical value $\beta_{\rm cr}$, we set $\beta = \beta_{\rm cr}$.  The mixing timescale is then calculated as 
\begin{equation}
    \rm{log}\,  \tau_{\rm{mix}}^{i} = \rm{log}\, \tau_{\rm{mix}}^{i+1} - \, \beta_{\rm cr} \, (\rm{log} \, P_i - \rm{log} \, P_{i+1}),
    \label{eq:tau_mix_overshoot}
\end{equation}
where $\tau_{\rm{mix}}^{i}$ is the mixing timescale at layer $i$ of the atmosphere, $\tau_{\rm{mix}}^{i+1}$ is the mixing timescale at layer $i+1$ of the atmosphere, which is one layer deeper than $i$, $P_{i, i+1}$ are the atmospheric pressures at layer $i$ and layer $i+1$ respectively. The parameterisation in Equation~\ref{eq:tau_mix_overshoot} has its origin in the numerical simulations of surface convection in late M-dwarfs by \citet{Ludwig2002}, as described in \citet{Woitke2004}. In this work we set $\beta_{\mathrm{cr}}$ = 2.2 following \citet{Ludwig2002}.

All the chemical surface reactions considered by \texttt{DRIFT} are listed in Table~\ref{tab:chemreak} in Appendix~\ref{ap:surfacereact}. 

\section{Atmosphere modelling with \texttt{MARCS}}
\label{sec:marcs}
\texttt{MARCS} is a one-dimensional, stratified, cloud-free, radiative-convective equilibrium atmospheric model in local thermodynamic equilibrium (LTE). \texttt{MARCS} models are computed with a \citet{Feautrier1964} type method, over a Rosseland optical depth scale and solved iteratively using a Newton-Raphson procedure \citep[e.g.][]{Nordlund1974, Gustafsson1975}{}{}. The radiative transfer scheme in \texttt{MARCS} uses a method of the type given by \citet{Rybicki1971}, as described in \citet{Gustafsson1972}. High-order angular dependencies are eliminated using the "variable Eddington factor"  technique \citep{Auer1970} as described in \citet{Cannon1973}. Convection is modelled with mixing length theory as described in \citet{Nordlund1974} and \citet{Gustafsson2008}. In this work, we use the most recent version of \texttt{MARCS}, i.e. \texttt{MSG} \citep{MSGpaper}, which is able to model plane-parallel cloud-free substellar atmospheres down to the effective temperatures of the coldest T-dwarfs ($\approx 300\,$K).

\subsection{Gas-phase equilibrium chemistry}
In \citet{MSGpaper}, \texttt{MARCS} was updated to use \texttt{GGchem} \citep{Woitke2018}, a thermochemical equilibrium code which is used to compute the gas-phase equilibrium chemistry in \texttt{MARCS}. \texttt{GGchem} functions by minimising the total Gibbs free energy, and is applicable across a wide temperature range, from 100\,K up to 6000\,K. \texttt{GGchem} computes the gas-phase equilibrium chemistry in each atmospheric layer from the local temperature, the gas pressure and the gas-phase element abundances. The cloud formation process depletes the local gas-phase element abundances and is taken into account explicitly through element conservation in the kinetic treatment of the cloud formation in \texttt{DRIFT}. The local gas-phase composition in each atmospheric layer is determined by \texttt{GGchem} using the cloud-depleted element abundances from the kinetic model. This is in contrast to models which use equilibrium condensation, or ``rainout" chemistry, where elements are removed from the gas-phase until the partial pressure of condensible species no longer exceeds the equilibrium vapour pressure.

The models presented here include 20 elements (H, He, C, N, O, Ne, Na, Mg, Al, Si, S, K, Ca, Cr, Fe, Ni, Li, Cl, Ti and V), each capable of existing as neutral atoms or singly charged ions. Following this choice of elements, \texttt{GGchem} considers 334 molecules, molecular ions and cations in the gas-phase equilibrium chemistry computations. Details on the thermochemical data and the methods used by \texttt{GGchem} can be found in \citet{Woitke2018}. 

\subsection{Gas and continuum opacities}

We include the same continuum opacities as in \citet{Juncher2017}. We compute the continuum absorption for 12 ions, electron scattering and Rayleigh scattering by H$_2$. The references for the continuum opacities data are listed in Table~\ref{tab:continuum} in Appendix~\ref{ap:opac_data}. In this work, we include the line opacities of 34 molecules and 2 atoms (Na and K). We sample the line opacities using the Opacity Sampling (OS) method as described in, for example, \citet{Jorgensen1992_OS}. All the models presented here use a sampling density of $R = \lambda/\Delta \lambda = 5000$. Specifically, the \texttt{TauREx} \citep{Taurex} OS files from the ExoMol database as compiled by \citet{Chubb2021} were modified by \citet{MSGpaper} as necessary to be read and used by \texttt{MSG}. All the references of the line lists of the molecules and atoms considered are presented in Table~\ref{tab:mol_opac} in Appendix~\ref{ap:opac_data}.

\section{The \texttt{MSG} model algorithm for cloudy substellar atmospheres}
\label{sec:sw_marcs}
\begin{figure*}[t]
\centering
    \includegraphics[width=\linewidth]{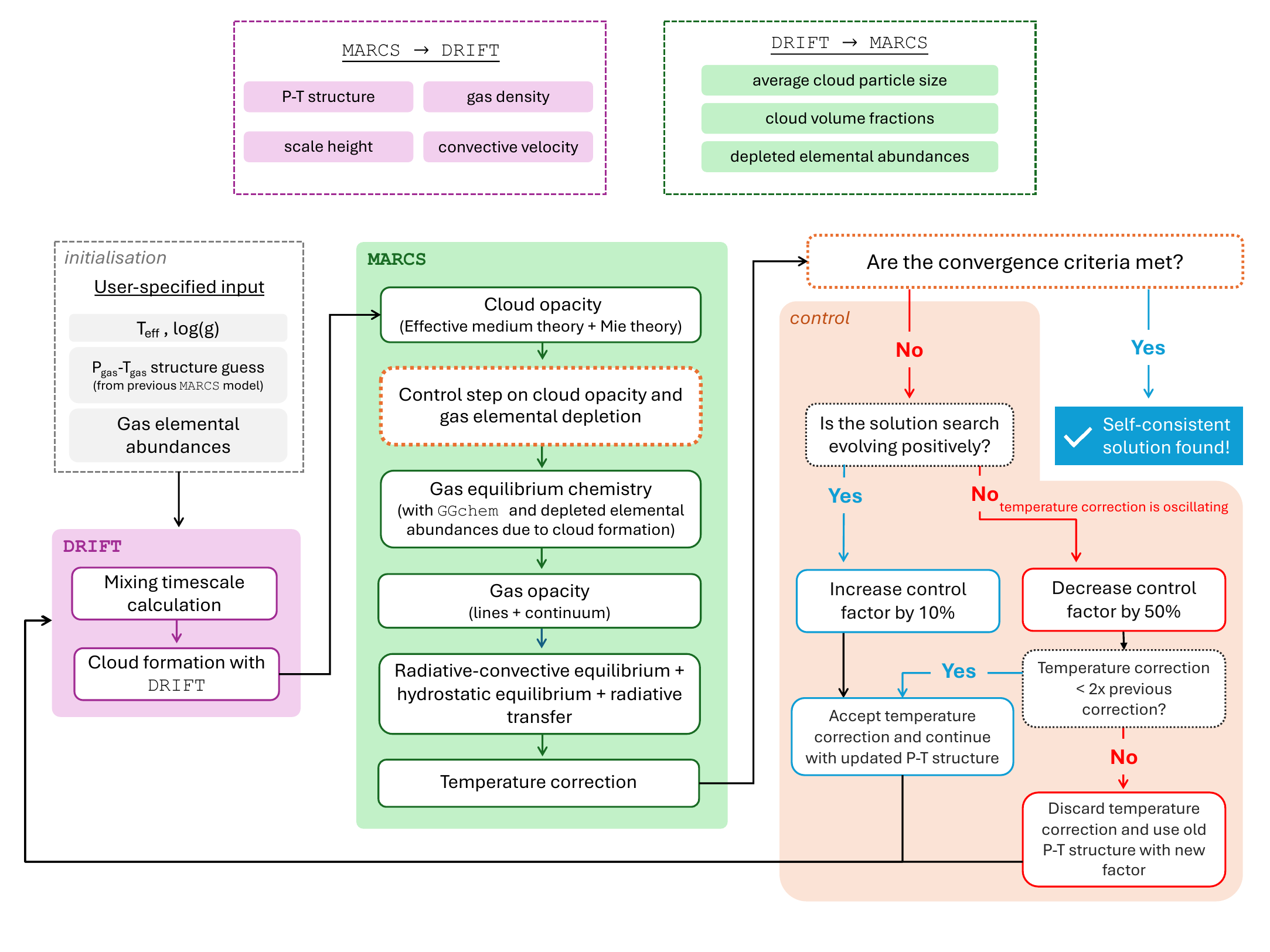}
    \caption{Diagram of the \texttt{MSG} model algorithm for cloudy substellar atmospheres. The boxes with a dashed outline indicate parameters that are inputs to the model. The boxes with a dotted outline indicate control processes within the workflow. For a detailed explanation of the cloud formation process, see section~\ref{sec:SW}. For a detailed explanation of the control process, see section~\ref{sec:control}. For the convergence criteria considered under the control process see section~\ref{sec:convergence_criteria}}.
    \label{fig:drift_marcs}
\end{figure*}
In the \texttt{MSG} models, we have coupled \texttt{MARCS} to \texttt{DRIFT} in a self-consistent manner to study the effects of microphysical cloud formation in substellar atmospheres. This implies that the radiative transfer scheme accounts for the cloud radiative feedback. The cloud radiative effect is added to the radiative transfer by considering the cloud's opacity contribution and through the change in the local gas opacity due to the depletion of cloud-forming elements. 

We run \texttt{MARCS} and \texttt{DRIFT} iteratively until we find a converged solution (the convergence criteria are defined in section~\ref{sec:convergence_criteria}). Fig.~\ref{fig:drift_marcs} shows a diagram of the \texttt{MSG} workflow. To run \texttt{MARCS}, the user is required to provide an effective temperature of the object ($T_{\rm{eff}}$), the gravitational acceleration of the object ($\log(g)$), the gas-phase element abundances, and an initial guess of the pressure-temperature structure of the atmosphere (typically from a previous \texttt{MARCS} model). The closer the initial pressure-temperature structure is to the solution, the faster the model converges. When running a grid of \texttt{MSG} models, it is best to start from a cloud-free model at a high $T_{\rm{eff}}$ ($\geq$ \,2500\,K), where we expect fewer clouds to form. This is because the clouds have a blanketing effect and heat up the atmosphere considerably compared to cloud-free models. As the $T_{\rm{eff}}$ decreases, the pressure-temperature structure of the cloudy atmosphere diverges more and more from that of a cloud-free atmosphere at the same $T_{\rm{eff}}$ and $\log(g)$. 

\texttt{DRIFT} requires as inputs from \texttt{MARCS} the pressure-temperature structure, the gas density, the scale height and the convective velocity, all as a function of atmospheric height. When starting a new run of \texttt{MSG}, these inputs are taken from a previous cloudy \texttt{MSG} model at a similar $T_{\rm{eff}}$ and $\log(g)$. The initial gas element abundances are set by the user and must be consistent between the two models. In \texttt{DRIFT}, the initial abundances are always the abundances before any element depletion by cloud formation has occurred. The computation of the mixing timescale (Equation~\ref{eq:tau_mix}) is done within the \texttt{DRIFT} framework. \texttt{DRIFT} then kinetically models the cloud formation considering the processes described in section~\ref{sec:SW}. The numerical methods are described in detail in \citet{Woitke2004}.

Once \texttt{DRIFT} has computed the cloud structure, we run \texttt{MARCS} taking as inputs from \texttt{DRIFT} the average cloud particle size, the cloud particle condensate volume fractions, and the depleted gas-phase element abundances, all as a function of atmospheric height. The cloud opacity is calculated using effective medium theory (EMT) and Mie theory (see section~\ref{sec:cloud_opac}). The change in the cloud opacity and in the gas-phase element abundances between two consecutive \texttt{MSG} iterations is controlled as described in the coming section~\ref{sec:control}. \texttt{MARCS} then proceeds to use \texttt{GGchem} to compute the gas-phase equilibrium chemistry. The gas opacities are calculated after obtaining the number densities for each gas-phase species from \texttt{GGchem}. The radiative-transfer equation is then solved using a \citet{Feautrier1964} type method, assuming radiative-convective equilibrium and hydrostatic equilibrium. The numerical methods are described in detail in \citet{Nordlund1974} and \citet{Gustafsson1975}. Finally, we obtain a temperature correction that is applied to the old $P_{\mathrm{gas}}-T_{\mathrm{gas}}$ structure to produce the new atmospheric structure. We check if the solution meets the convergence criteria. If the convergence criteria are not met, we enter a control process to handle the cloud opacity and the depleted gas-phase element abundances in order to avoid oscillations in the temperature corrections and reach convergence. This control process is explained in detail in section~\ref{sec:control}. The cycle described above is repeated until a converged solution is found.

\subsection{Convergence criteria}
\label{sec:convergence_criteria}
The convergence criteria for a \texttt{MSG} cloudy model are the following: 
\begin{enumerate}
    \item The change in the $P_{\mathrm{gas}}-T_{\mathrm{gas}}$ structure between the previous and current iteration must be smaller than 5\,K;
    \item The relative difference in the cloud opacity and the gas-phase element abundances between the current and the previous iteration must be smaller than 10\%;
    \item The relative difference in the wavelength-dependent emergent flux between the current and previous iteration must be smaller than 1\%. Generally, when the wavelength-dependent emergent flux has converged, $|F^{\lambda}_i - F^{\lambda}_{i-1}|/F^{\lambda}_i < 1\%$, all the other convergence criteria have also been met.
\end{enumerate}

\subsection{Cloud opacity}
\label{sec:cloud_opac}
We consider the cloud particles to be mixed-material, well-mixed, spherical, and compact. As presented in \citet{Lee2016}, we calculate the cloud opacity with spherical particle Mie theory \citep{Mie1908} combined with effective medium theory (EMT).

The effective optical constants for the material mixtures are calculated with EMT. We generally use the numerical Bruggeman method \citep{Bruggeman1935} except for rare cases where we find non-convergence and therefore use the analytic Landau-Lifshitz-Looyenga [LLL] method \citep{Looyenga1965} (see section 2.4. of \citet{Lee2016} for more details). The cloud particle extinction and scattering coefficients are computed with Mie theory using the routine developed by \citet{Wolf2004}, which is based on the widely used \citet{Bohren1983} routine.

We use the data tables compiled by \citet{Kitzmann2018} for the cloud particle optical constants. Table~\ref{tab:dust} in Appendix~\ref{ap:dust} lists all the references for the cloud particle optical constants.

\subsection{Controlling the cloud opacity and the depleted gas element abundances}
\label{sec:control}

One of the biggest challenges in modelling cloudy atmospheres self-consistently is handling the significant radiative feedback of the cloud. At effective temperatures of less than 2200\,K, if we start a \texttt{MSG} run from a cloud-free model, the atmosphere is significantly heated up at the second iteration due to the cloud radiative feedback. When we run the cloud formation model with this heated-up atmosphere as an input, much of the condensed material is no longer thermally stable because the atmosphere is significantly hotter, and therefore evaporates. This leads to an atmosphere that is now significantly cooler than the one obtained in the previous iteration. The solution may thus enter an oscillating cycle without ever converging. 

We introduce a control factor $f$ in our modelling to avoid these oscillations. The factor can take a value between 0 and 1, and it controls the change in cloud opacity and in element abundances between two iterations. We define the cloud opacity $\kappa^{\rm{cloud}}$ at iteration $j$ as
\begin{equation}
\kappa^{\rm{cloud}}_j = f\, \kappa^{\rm{cloud}}_{j/2} + (1-f)\,\kappa^{\rm{cloud}}_{j-1},
\end{equation}
where $\kappa^{\rm{cloud}}_{j-1}$ is the cloud opacity used in the previous iteration and $\kappa^{\rm{cloud}}_{j/2}$ is the true cloud opacity calculated after the \texttt{DRIFT} run (first step within \texttt{MARCS} as shown in the diagram in Fig.~\ref{fig:drift_marcs}). We update the gas-phase element abundances in the same manner. The gas-phase element abundance $\epsilon_i$ of element $i$ at iteration $j$ is therefore given by 
\begin{equation}
\epsilon^{j}_i = f\, \epsilon^{j/2}_{i} + (1-f)\,\epsilon^{j-1}_{i},
\end{equation}
where $\epsilon^{j-1}_{i}$ is the element abundance used in the previous iteration and $\epsilon^{j/2}_{i}$ is the true element abundance computed after the \texttt{DRIFT} run.

The factor $f$ is updated depending on the evolution of the solution search. The solution search is evolving positively if the atmosphere is consecutively heating up or cooling down. This is checked by considering the ratio between the maximum temperature corrections obtained in the current and previous iterations, i.e. 
\begin{equation}
    R = \frac{\Delta T^{\mathrm{max}}_j}{\Delta T^{\mathrm{max}}_{j-1}},
\end{equation}
where $\Delta T^{\mathrm{max}}_{j,j-1}$ are the maximum temperature corrections in the current and previous iterations respectively. 
If the ratio $R$ is positive, the atmosphere has either heated up consecutively between the two iterations or cooled down consecutively between the two iterations. This means the solution is evolving positively and we accept the temperature correction and increase the factor $f$ by 10\%. If the ratio $R$ is negative, we are within an oscillation and decrease $f$ by 50\%. These percentages were chosen after we performed testing on a toy model and concluded they tend to minimise the number of iterations needed for convergence. If the current temperature correction exceeds double the previous temperature correction, we discard the current temperature correction and re-run the model with the new $f$. Otherwise, we accept the temperature correction but still use the new $f$ in the next iteration. These steps are illustrated in the area denominated ``control'' in the diagram shown in Fig.~\ref{fig:drift_marcs}.

We note the maximum value $f$ can have is 1 and $f$ can never be 0. At effective temperatures above 1800\,K, we start the model run with $f=1$. However, at lower effective temperatures we have found it favourable to start with a lower value of $f$. We typically use $f=0.1$. For the 0$^{th}$ iteration, $j=0$, $\kappa^{\rm{cloud}}_{-1}$ is the cloud opacity used in the last iteration of the input model. If the input model is cloud-free then $\kappa^{\rm{cloud}}_{-1} = 0$ and $f$ is set to 1. The cloud opacity is converged if $| \kappa^{\rm{cloud}}_{i/2} - \kappa^{\rm{cloud}}_{i}|/\kappa^{\rm{cloud}}_{i/2}$ is less than 10\%. The same principles apply to the control of each of the gas-phase element abundances. 
%__________________________________________________________________
\section{Results}
\label{sec:results}
The base grid of models consists of 11 models at effective temperatures between 2500\,K and 1500\,K, in steps of 100\,K, at $\log(g)=4.0$, with undepleted solar element abundances and C/O ratio, and TiO$_2$ CCN. In sections~\ref{sec:pt}, ~\ref{sec:clouds} and ~\ref{sec:spectra} we present $P_{\mathrm{gas}}-T_{\mathrm{gas}}$ profiles, cloud structures and properties, and model spectra for some selected models. In section~\ref{sec:nucleation}, we explore the effect of changing the CCN species to SiO. In section~\ref{sec:mixing} we explore the effect of scaling up the mixing timescale, making the mixing less efficient.

\subsection{$P_{\mathrm{gas}}-T_{\mathrm{gas}}$ profiles}
\label{sec:pt}
\begin{figure}[t]
   \centering
    \resizebox{\hsize}{!}{\includegraphics{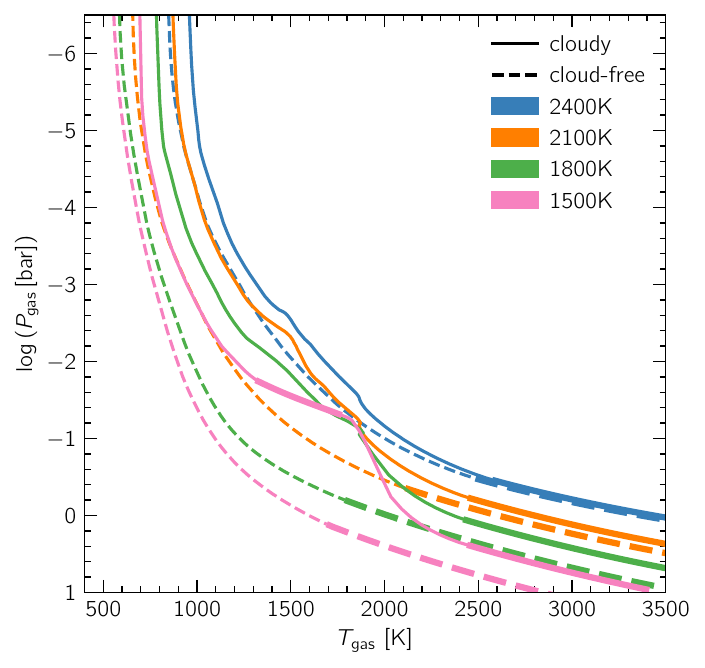}}
    \caption{Pressure-temperature profiles for cloud-free \texttt{MSG} models (dashed curves) and cloudy \texttt{MSG} models with TiO$_2$ nucleation (solid curves), at different effective temperatures and $\log(g)=4.0$. Convective zones are plotted with thicker lines while radiative zones are plotted with thinner lines. We note the cloudy profile at 1500\,K has a detached convective zone.}
    \label{fig:pt_profiles}
\end{figure}

Fig.~\ref{fig:pt_profiles} shows the $P_{\mathrm{gas}}-T_{\mathrm{gas}}$ profiles of \texttt{MSG} cloudy models (solid curves) and \texttt{MSG} cloud-free models (dashed curves), at the effective temperatures of 2400\,K, 2100\,K, 1800\,K and 1500\,K, and $\log(g)=4.0$. The cloudy models are consistently warmer than the cloud-free models at the same effective temperature, indicating the clouds have a blanketing effect over the atmosphere. This effect is generally seen in L dwarf models \citep[e.g.][]{Morley2024}{}{}. 
The radiative regions of the atmosphere are plotted with a thinner line width, while the convective regions are plotted with a thicker line width. 

In the cloudy models, a detached convective zone emerges at $T_{\rm{eff}}$ $\leq$ 1600\,K  due to the increasing cloud opacity. This is also seen in other L dwarf model grids \citep[e.g.][]{Marley1996,Burgasser02, Burrows06, Saumon2008, Witte2011,Malik2019, Morley2024}{}{}. In section ~\ref{sec:conv} we discuss the implications of detached convective zones in more detail. 

% The exact pressure of the L changes between models due to the \texttt{MARCS} Rosseland tau grid (see Sect.~\ref{sec:marcs}). As the cloud properties and, consequently, the detailed effect of the radiative feedback changes, the pressure corresponding to each Rosseland tau value also changes.

\subsection{Cloud structure and properties}
\label{sec:clouds}
\begin{figure*}[t]
\centering
%\begin{minipage}[b]{0.33\linewidth}
     \includegraphics[width=0.33\linewidth, valign=t]{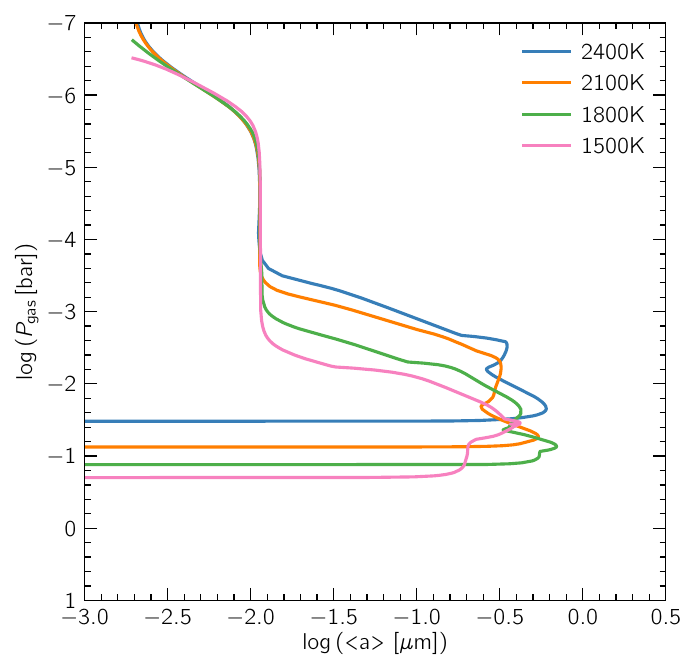}
%\end{minipage}
%\begin{minipage}[b]{0.33\linewidth}
     \includegraphics[width=0.33\linewidth, valign=t]{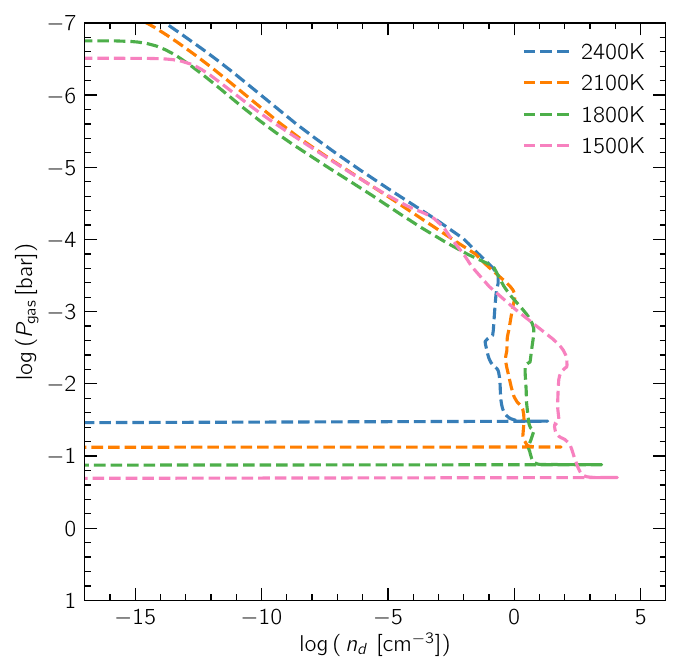}
%\end{minipage}
%\begin{minipage}[b]{0.33\linewidth}
     \includegraphics[width=0.33\linewidth, valign=t]{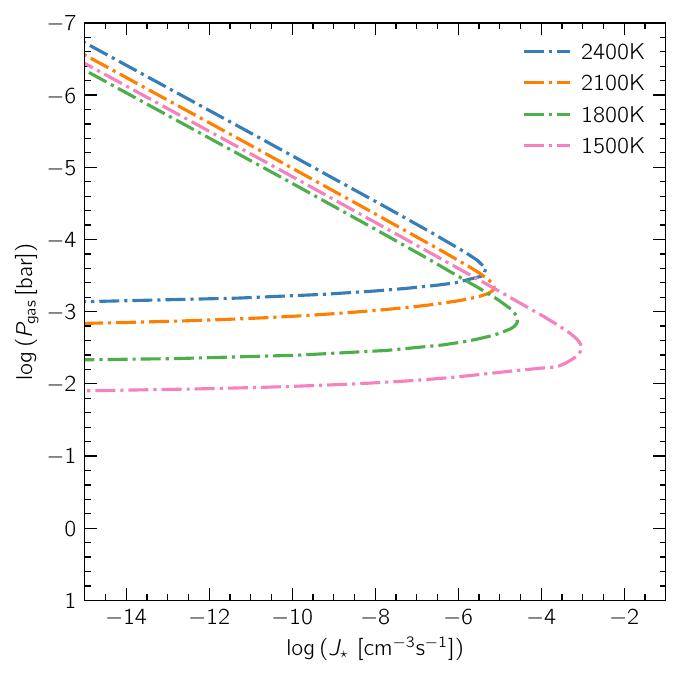}
%\end{minipage}
     \caption{The average cloud particle size $\langle a \rangle$ (left), the cloud particle number density $n_d$ (middle) and the nucleation rate $J_\star$ (right) along the atmosphere for models with TiO$_2$ nucleation at $T_{\rm{eff}}$ $=$ 2400\,K, 2100\,K, 1800\,K and 1500\,K and $\log(g)=4.0$. The corresponding $P_{\mathrm{gas}}-T_{\mathrm{gas}}$ profiles are shown in Fig.~\ref{fig:pt_profiles}.}
     \label{fig:cloud_size_nd_TiO2}
\end{figure*}
\begin{figure*}[t]
\centering
%\begin{minipage}[b]{0.33\linewidth}
   \includegraphics[width=0.33\linewidth, valign=b]{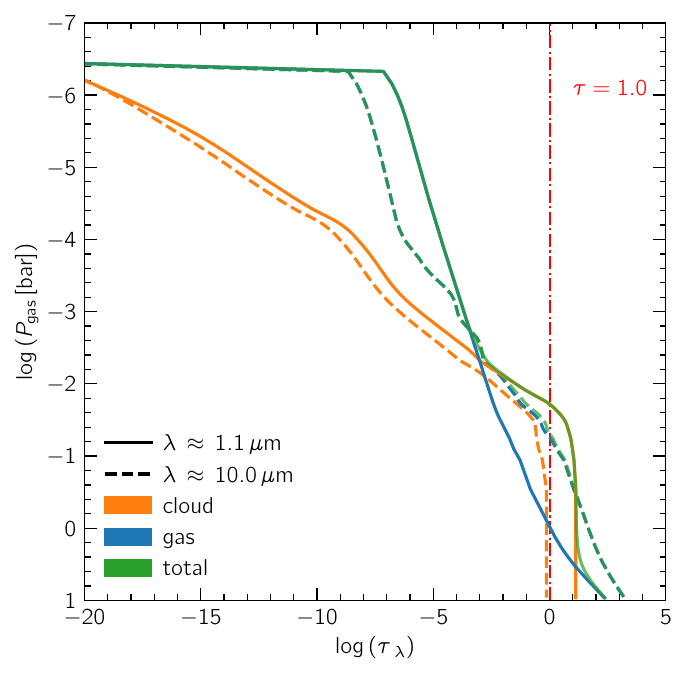}
%\end{minipage}
%\begin{minipage}[b]{0.33\linewidth}
   \includegraphics[width=0.33\linewidth, valign=b]{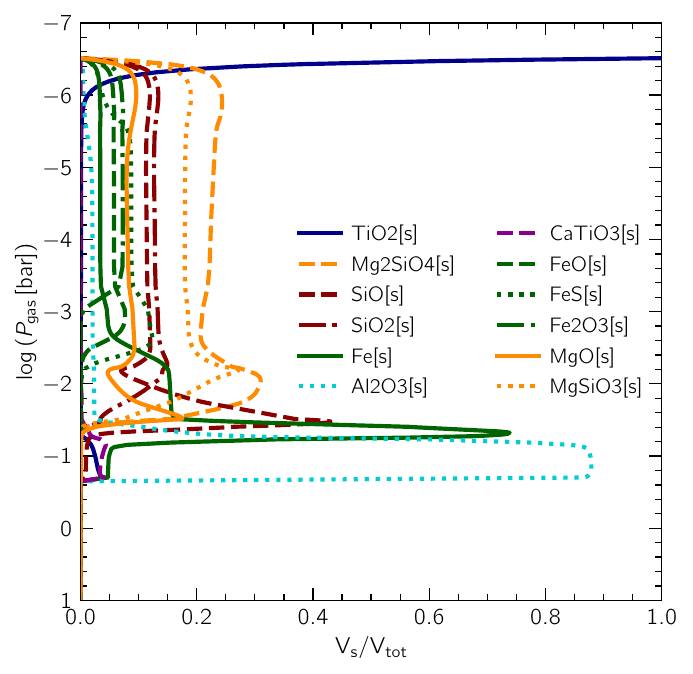}
%\end{minipage}
%begin{minipage}[b]{0.33\linewidth}
   \includegraphics[width=0.33\linewidth, valign=b]{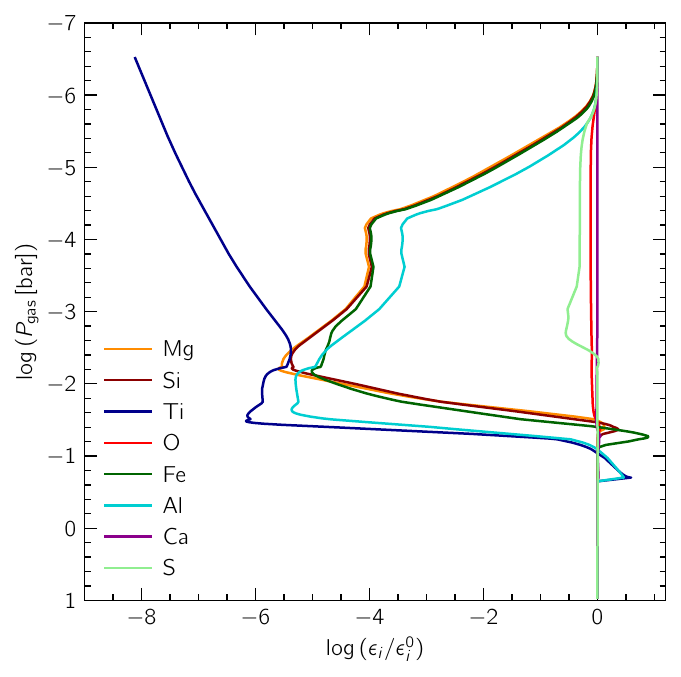}
%\end{minipage}
     \caption{\textbf{Left:} The optical depth of the cloud (orange), gas (blue), and the total (green) throughout the atmosphere at $\lambda \approx 1.1 \,\mu$m (solid curves) and $\lambda \approx 10.0 \,\mu$m (dashed curves), for a \texttt{MSG} model with TiO$_2$ nucleation, at 1500\,K and $\log(g)=4.0$. The $\tau=1.0$ level is shown for reference (dot-dash red curve). \textbf{Middle and right:} Composition of the cloud particles in units of volume fractions $V_s/V_{\mathrm{tot}}$ (middle) and the relative gas-phase element depletions $\epsilon_i/ \epsilon_i^0$ (right) for the same model as in the left panel. }
     \label{fig:cloud_comp_TiO2}
\end{figure*}

Fig.~\ref{fig:cloud_size_nd_TiO2} shows the average cloud particle size (left), the cloud particle number density (middle) and the nucleation rate (right) in the atmosphere for the \texttt{MSG} models shown in Fig.~\ref{fig:pt_profiles}. Fig.~\ref{fig:cloud_comp_TiO2} left panel shows the optical depth of the cloud and gas along the atmosphere at the wavelengths of 1.1\,$\mu$m and 10.0\,$\mu$m, for the model at 1500\,K.Fig.~\ref{fig:cloud_comp_TiO2} middle and right panels show the cloud particle composition and the relative gas-phase elemental depletion along the atmosphere also for the model at 1500\,K.

We can identify four different cloud stages where different processes take place and dominate the cloud structure. The first region we identify is the region of the \textit{nucleation} and \textit{first growth} stage, visible in Fig.~\ref{fig:cloud_size_nd_TiO2} (left) between $\sim 10^{-7}$ to $10^{-5}$ bar. As the cloud particles fall inwards, the increasing gas density and the element replenishment allow for the collision rate between the gas and the dust particles to increase, allowing for a growing number of surface reactions. This leads to the cloud particles growing in size and the cloud-forming elements depleting significantly (see Fig.~\ref{fig:cloud_comp_TiO2} right). Particularly, we see Ti highly depleted at the very top of the atmosphere (TOA) due to the \textit{nucleation} process. Although O is also depleted, this is not as visible because O is highly abundant compared to Ti. Throughout this stage, the nucleation rate continues to increase, however the growth process is dominant on the average cloud particle size. 

We then reach the \textit{drift} stage, between $\sim 10^{-5}$ to $10^{-3}$ bar (dependent on the effective temperature). During this stage, the nucleation rate is increasing. This results in the formation of many new, small cloud particles, efficiently consuming elements from the gas-phase. Consequently, the available material for the growth of cloud particles is reduced, and the net growth rate of the particles decreases. The creation of new particles and the reduction in the net growth rate ultimately act to keep the average particle size constant over this pressure region. The maximum gas-phase depletion is reached at about the same pressure at which the nucleation rate peaks (Fig.~\ref{fig:cloud_comp_TiO2} right).

When the nucleation rate drops, the increasing grain size is no longer balanced by the formation of new CCN, and therefore we reach the \textit{second growth} stage,  between $\sim 10^{-3}$ to $10^{-1}$ bar (dependent on the effective temperature). At this point, due to backwarming by the cloud particles, the temperature increases rapidly with increasing gas pressure (see Fig.~\ref{fig:pt_profiles}). The silicate species and magnesium oxide are the first to react to this. Over a small pressure interval, dissociation begins to be the favoured chemical path. We enter the stage of \textit{evaporation}. The average particle size drops slightly, which creates a local maximum in the average cloud particle size (Fig.~\ref{fig:cloud_size_nd_TiO2} left ). For example, for the model at 1500\,K, we can see from Fig.~\ref{fig:cloud_comp_TiO2} (middle) that the first local maximum in cloud particle size occurs just before the cloud volume fraction of SiO[s], SiO$_2$[s], Mg$_2$SiO$_4$[s], MgSiO$_3$[s] and MgO[s] drop to zero (between 0.01 and 0.1 bar). This is also seen in Fig.~\ref{fig:cloud_comp_TiO2} (middle), where the relative element abundance of Si and Mg are replenished due to the evaporation. 

Once the silicates evaporate, Fe[s] and AlO$_3$[s] dominate the growth process. Fe[s] quickly becomes the species with the largest cloud volume fraction, increasing the cloud opacity in the optical and near-infrared (NIR). Due to the increased cloud opacity, the cloud particles heat extremely fast, and Fe[s] quickly evaporates, followed by AlO$_3$[s], CaTiO$_3$[s] and TiO$_2$[s]. A local maximum in average cloud particle size is also visible for the evaporation of  Fe[s] and AlO$_3$[s].

The cloud ends when all the cloud species have evaporated. Within the evaporation regions of individual solids, cloud-forming elements can be enriched due to their rain out in cloud particles (Fig.~\ref{fig:cloud_comp_TiO2} middle). 

The maximum average cloud particle size is similar at all effective temperatures modelled. However, at the same pressure, the average cloud particle size is smaller for decreasing effective temperature. There is a shift with effective temperature in the pressure at which the \textit{second growth} starts. The \textit{second growth} starts at higher pressures for smaller effective temperatures (Fig.~\ref{fig:cloud_size_nd_TiO2} left). This is expected as the nucleation rate peaks at higher pressures with decreasing effective temperature (Fig.~\ref{fig:cloud_size_nd_TiO2} right). The cloud particle number density increases slightly with effective temperature due to the reduction in the average cloud particle size (Fig.~\ref{fig:cloud_size_nd_TiO2} middle). 

Regarding the optical depth, in Fig.~\ref{fig:cloud_comp_TiO2} (left) we see that around 1.1\,$\mu$m the cloud becomes optically thick around 0.4\,bar. However, in the MIR around 10.0\,$\mu$m where we expect the silicate feature in the observed spectra, the cloud never becomes optically thick. We discuss the optical properties of the cloud in more detail in section~\ref{sec:cloud_opt_properties}.
\subsection{Synthetic spectra}
\label{sec:spectra}

\begin{figure*}[t]
\centering
     \includegraphics[width=0.48\linewidth, valign=t]{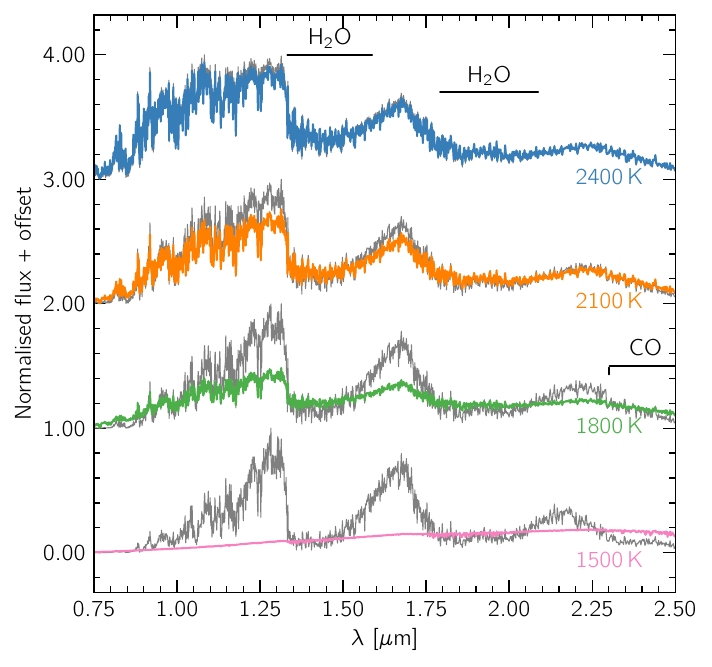}
%\begin{minipage}[b]{0.5\linewidth}
     \hspace{0.02\linewidth}
     \includegraphics[width=0.48\linewidth, valign=t]{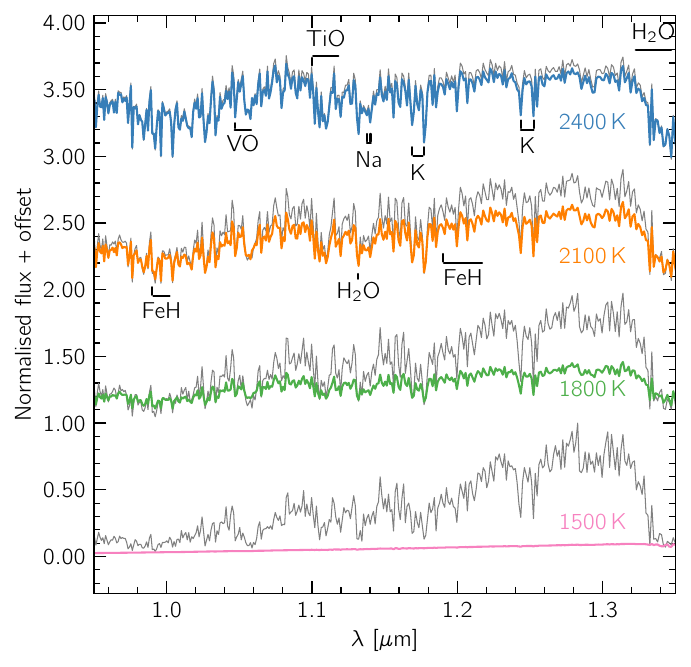}
%\end{minipage}
%\begin{minipage}[b]{0.5\linewidth}
     \includegraphics[width=0.48\linewidth, valign=t]{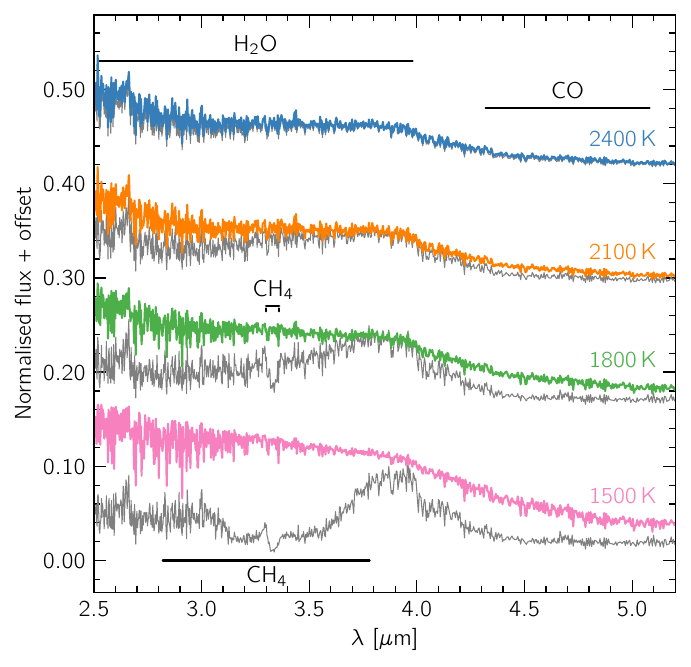}
     \hspace{0.03\linewidth}
%\end{minipage}
%\begin{minipage}[b]{0.5\linewidth}
     \includegraphics[width=0.48\linewidth, valign=t]{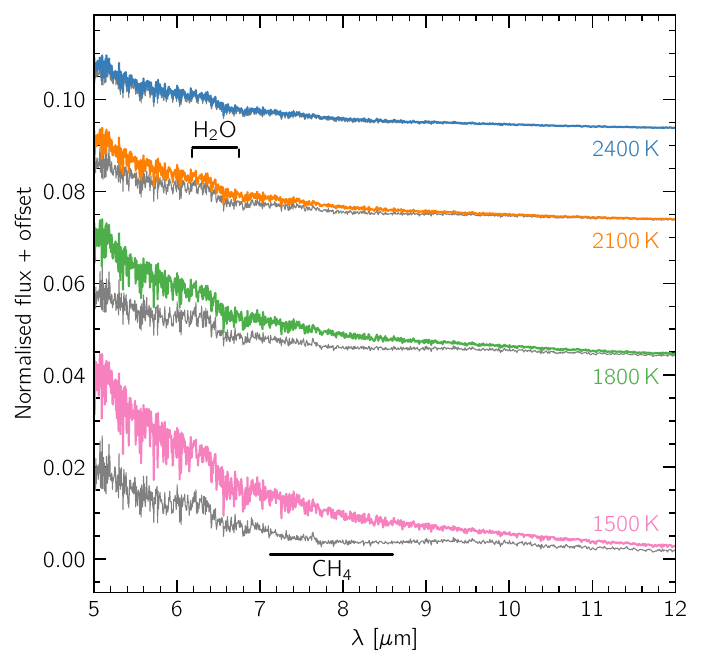}
%\end{minipage}

     \caption{Synthetic spectra of \texttt{MSG} models with TiO$_2$ nucleation at $T_{\rm{eff}}$ $=$ 2400\,K (blue), 2100\,K (orange), 1800\,K (green) and 1500\,K (pink), with $\log(g)=4.0$, and the respective cloud-free spectra at the same $T_{\rm{eff}}$ and $\log(g)$ in grey. The emergent fluxes are normalised with respect to the cloud-free \texttt{MSG} spectra, i.e. $F_{\mathrm{norm}}(\lambda) = F_{\mathrm{cloudy}}(\lambda) / F^{\mathrm{max}}_{\mathrm{cloud-free}}$, where $F_{\mathrm{norm}}$ is the normalised flux, $F_{\mathrm{cloudy}}$ the flux from the cloudy model, and $F^{\mathrm{max}}_{\mathrm{cloud-free}}$ the maximum flux from the cloud-free model within the full wavelength range considered ($\sim$ 0.4$\,\mu$m\,-\,20.0$\,\mu$m). An arbitrary offset is added for clarity. In the top-left plots we show the NIR range, in the top-right the Y and J bands, in the bottom-left the MIR and the bottom-right the thermal infrared. Important absorbers are respectively labelled.}
     \label{fig:spectra_TiO2}
\end{figure*}
Synthetic spectra allow us to compare models to observations. In this section, we show the spectra of the \texttt{MSG} models presented in Fig.~\ref{fig:pt_profiles} at selected wavelength ranges. All the spectra shown have been re-binned to have a resolution of $R = 1000$. The spectra inform us about the observable atmosphere of the object. The observable atmosphere of the object goes down to where the optical depth ($\tau$) is equal to unity. Anything below $\tau=1$ is not observable. 

\begin{figure*}[t]
\centering
\includegraphics[width=0.48\linewidth]{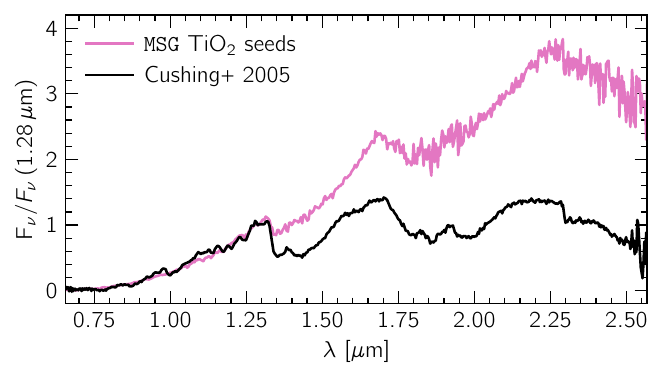}
\includegraphics[width=0.48\linewidth]{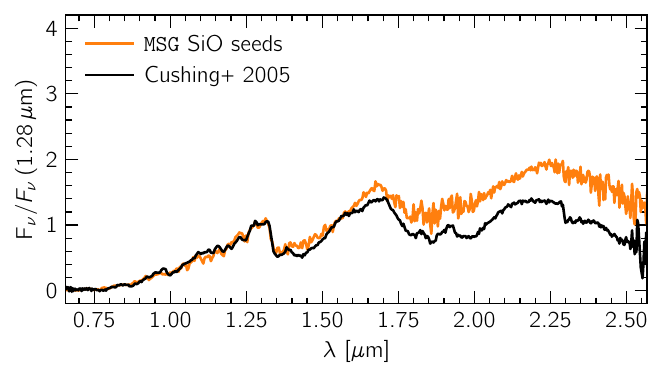}
\includegraphics[width=0.485\linewidth]{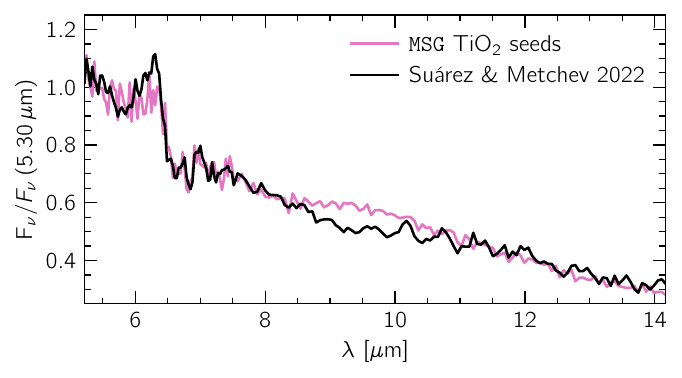}
\includegraphics[width=0.485\linewidth]{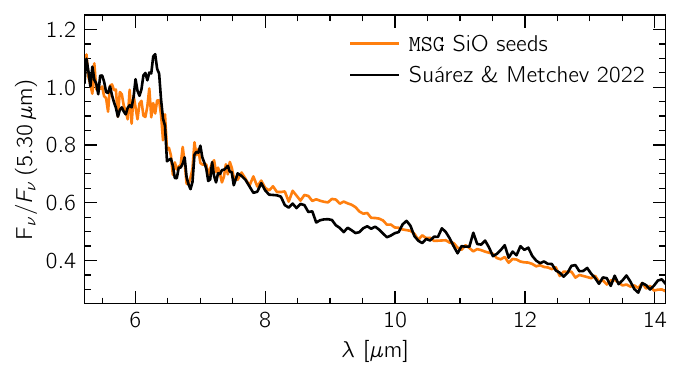}

\caption{Comparison between observed spectra of 2MASS 1507-1627 (L5) and \texttt{MSG} cloudy models. 2MASS 1507-1627 has an effective temperature $\sim$1600\,K \citep{Cushing2005} and a surface gravity of about log(g)$\approx$5.2 \citep{Filippazzo2015}. The observed data is shown in black. The NIR data (top panels) is from \citet{Cushing2005}, and the MIR data (bottom panels) is from \citet{Suarez2022}. In the left panels, the \texttt{MSG} model spectra shown in pink is for a model with \Teff \,= 1600\,K, log(g)=4.0 and \ce{TiO2} nucleation seeds. In the right panels, the \texttt{MSG} model spectra shown in orange is for a model with \Teff \,= 1600\,K, log(g)=4.0 and \ce{SiO} nucleation seeds. We note, unlike other figures in the paper, here we plot the frequency dependent flux ($F_\nu$) for a better visualisation of the silicate feature around 10\,$\mu$m in the MIR.}
\label{fig:obs_tio2}
\end{figure*}

Fig.~\ref{fig:spectra_TiO2} shows \texttt{MSG} cloudy and cloud-free spectra at different wavelength ranges, for the models shown in Fig.~\ref{fig:pt_profiles}. At the highest effective temperatures (2100\,K and 2400\,K in the plots), the cloudy spectra are similar to the cloud-free spectra. They are majorly defined by the absorption features of atoms Na and K, metal oxides such as TiO and VO, and the metal hydride FeH, in the optical and NIR (Fig.~\ref{fig:spectra_TiO2}, top right). In the MIR, H$_2$O and CO dominate the shape of the spectra (Fig.~\ref{fig:spectra_TiO2}, bottom left). 

Towards lower effective temperatures, the effects of the cloud opacity are significant in all wavelength bands. The spectra redden significantly in the NIR wavelengths, and the emergent flux is considerably reduced (Fig.~\ref{fig:spectra_TiO2}, top left). For example, although the abundances of Na and K are never depleted due to cloud formation, the absorption features of both these atoms shrink with decreasing effective temperature (Fig.~\ref{fig:spectra_TiO2} top right). At 1500\,K the Na and K features are no longer visible due to the cloud continuum. Another striking difference between the cloud-free and cloudy spectra is the effective temperature at which CH$_4$ emerges. For the cloud-free models, the CH$_4$ feature at $\sim 3.3\, \mu$m emerges at approximately 1900\,K. Due to the warming of the $P_{\mathrm{gas}}-T_{\mathrm{gas}}$ structure by the cloud radiative feedback, CH$_4$ never emerges as prominently in the spectra within the models shown in Fig.~\ref{fig:spectra_TiO2} (bottom right). \citet{Morley2024} see the exact same effect in their model grid.
%\newpage

In the models presented in this section, the cloud opacity has the extreme effect of making the spectra almost blackbody-like in the NIR for effective temperatures below 1600\,K. This is extremely different compared to observed spectra \citep[e.g.][]{Cushing2005, Cushing2008, Stephens2009, Suarez2022, Miles2023}{} where, although a cloud continuum is generally seen, the spectra are not blackbody-like, presenting plenty of absorption features. Our model spectra are also redder in the NIR than observations. In addition to this, the observed silicate feature of the clouds around 10.0\,$\mu$m \citep[e.g.][]{Cushing2006, Suarez2022, Miles2023} is missing. In Fig.~\ref{fig:obs_tio2} left panels, we compare one \texttt{MSG} model with \ce{TiO2} seeds to observations in the NIR and MIR of 2MASS 1507-1627 by \citet{Cushing2005} and \citet{Suarez2022} respectively. It is straightforward to see the discrepancies described above between our model grid and observations. Therefore, in the next two sections, we explore nucleation and mixing parameterisations to unravel the potential origin of our redder NIR spectra. Additionally, in section~\ref{sec:cloud_opt_properties} we explore the optical properties of the cloud to investigate why the silicate feature is missing.
\begin{figure*}[t]
    \includegraphics[width=0.34\linewidth]{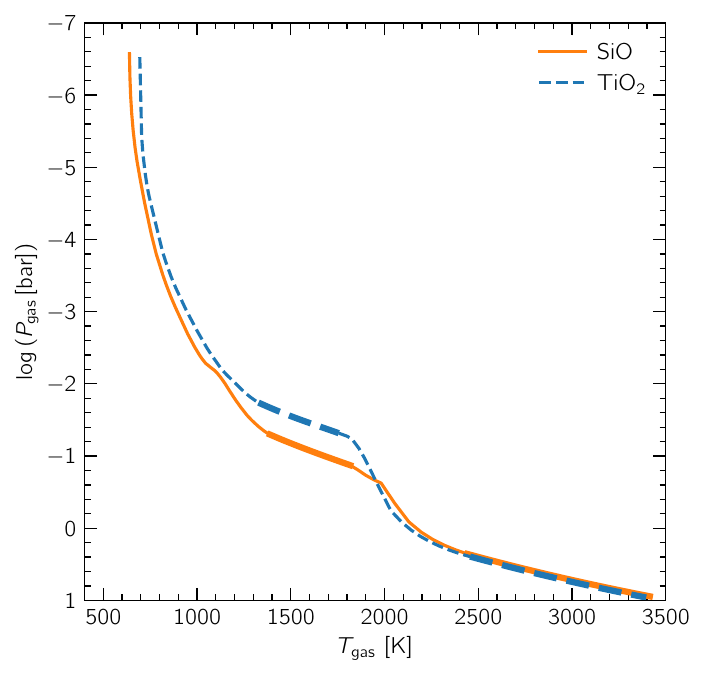}
    \hspace{-8.0pt}
    \includegraphics[width=0.33\linewidth, valign=b]{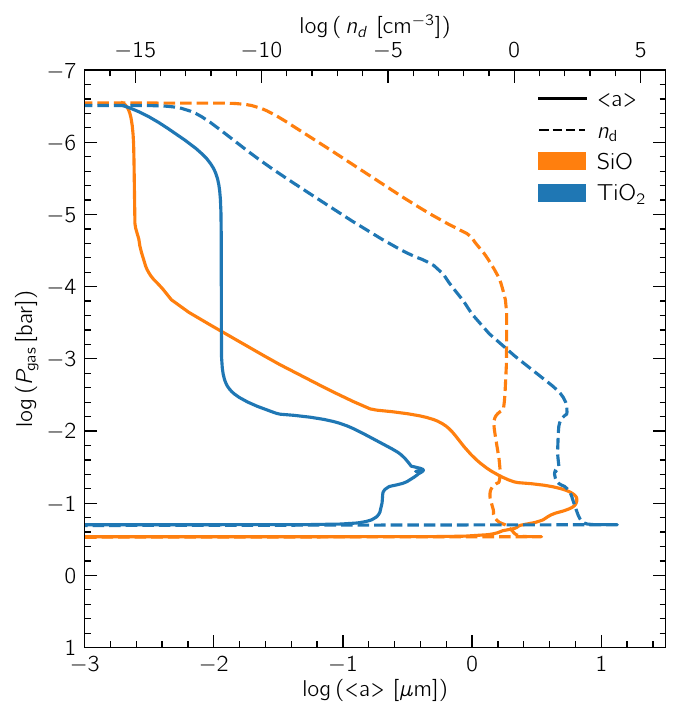}
    \includegraphics[width=0.33\linewidth, valign=b]{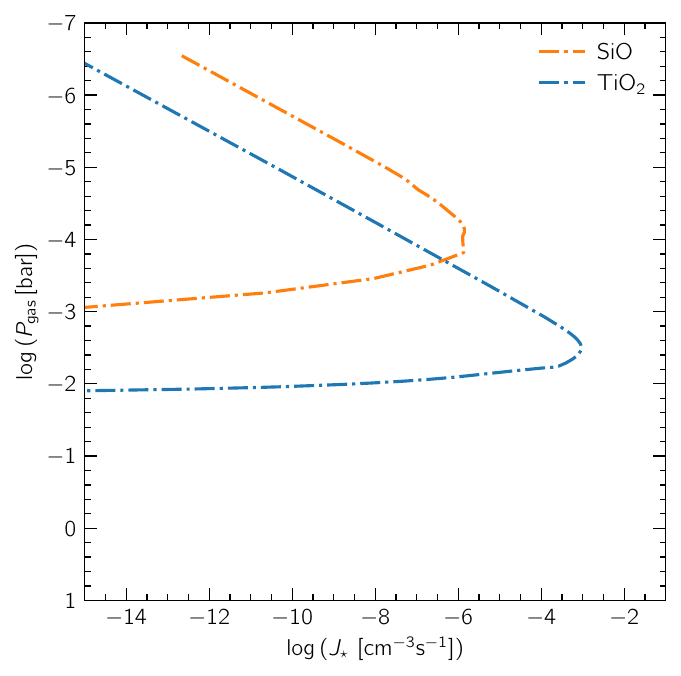}
    \caption{\textbf{Left}: Pressure-temperature profiles for \texttt{MSG} models at $T_{\rm{eff}}$$=$ 1500\,K and $\log(g)=4.0$ with \ce{TiO2} nucleation (blue dashed curve) and \ce{SiO} nucleation (orange solid curve). Convective zones are plotted with thicker lines while radiative zones are plotted with thinner lines. \textbf{Middle:} The average cloud particle size (solid curves) and the cloud particle number density (dashed curves) for the models shown on the left. \textbf{Right:} The nucleation rates for \texttt{MSG} models at $T_{\rm{eff}}$ $=$ 1500\,K and $\log(g)=4.0$ with \ce{TiO2} nucleation (blue curve) and \ce{SiO} nucleation (orange curve).}
    \label{fig:pt_SiO}
\end{figure*}
\begin{figure*}[t]
    \includegraphics[width=0.335\linewidth, valign=t]{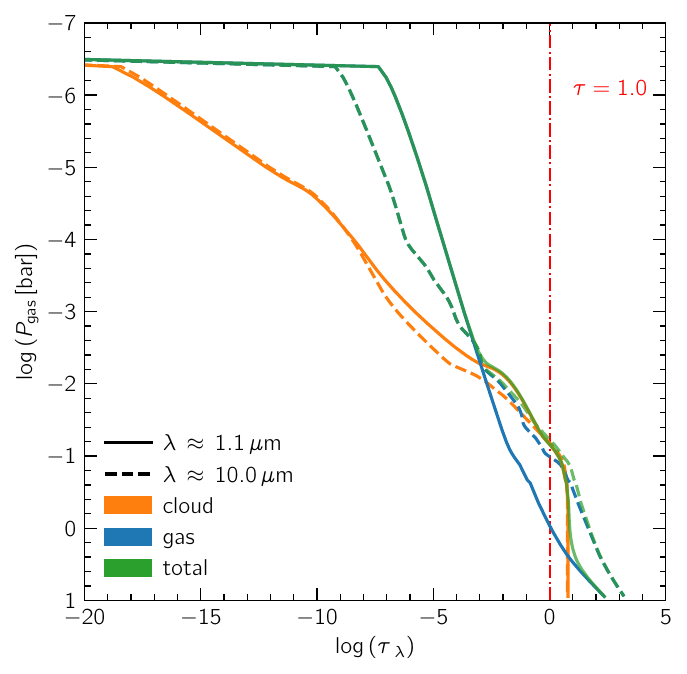}
    \hspace{-8.0pt}
    \includegraphics[width=0.335\linewidth, valign=t]{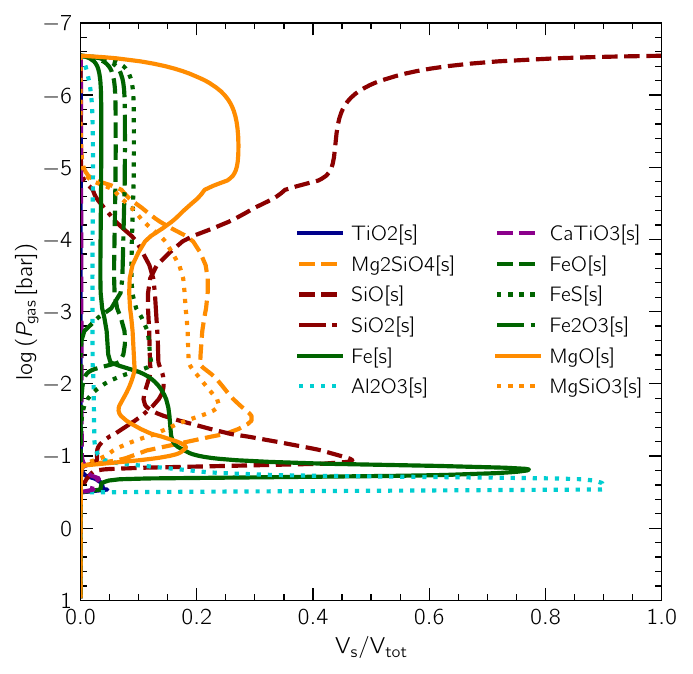}
    \hspace{-8.0pt}
    \includegraphics[width=0.335\linewidth, valign=t]{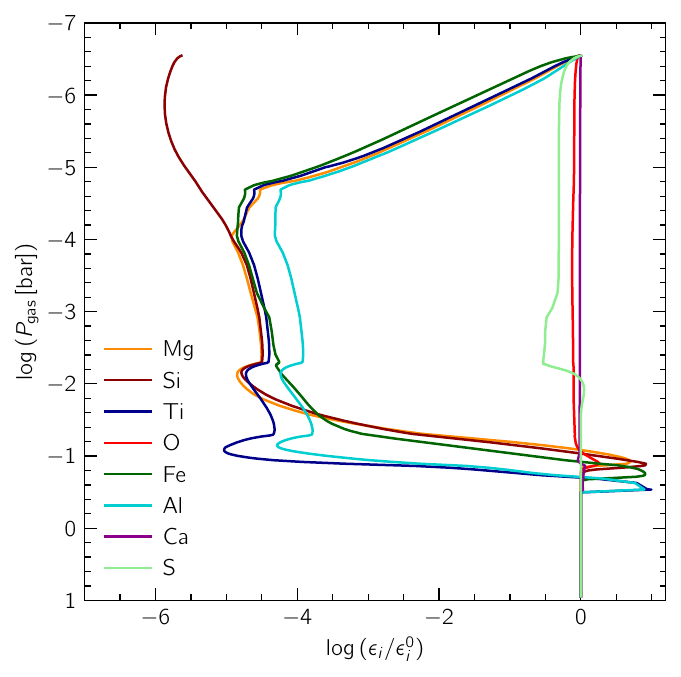}
    \caption{\textbf{Left:} The optical depth of the cloud (orange), gas (blue), and the total (green) throughout the atmosphere at $\lambda \approx 1.1 \,\mu$m (solid curves) and $\lambda \approx 10.0 \,\mu$m (dashed curves), for a \texttt{MSG} model with SiO nucleation, at 1500\,K and $\log(g)=4.0$. The $\tau=1.0$ level is shown for reference (dot-dash red curve). \textbf{Middle and right:} Composition of the cloud particles in units of volume fractions $V_s/V_{\mathrm{tot}}$ (middle) and the relative gas-phase element depletions $\epsilon_i/ \epsilon_i^0$ (right) for the same model as in the left panel.}
    \label{fig:cloud_comp_SiO}
\end{figure*}
\subsection{Models with SiO nucleation}
\label{sec:nucleation}
In this section, we investigate the effect of changing the CCN from TiO$_2$ to SiO in our self-consistent models. An in-depth investigation comparing \ce{TiO2} to SiO nucleation with the \texttt{DRIFT} model was previously conducted by \citet{Lee2015}. However, \citet{Lee2015} do this comparison by post-processing \texttt{DRIFT} on \texttt{DRIFT-PHOENIX} models \citep{Witte2009}. Here, we make this comparison using our full self-consistent algorithm and investigate the impact a different CCN has on the atmospheric $P_{\mathrm{gas}}-T_{\mathrm{gas}}$ structure, the cloud structure and the observables (spectra).

\begin{figure*}[t]
\begin{minipage}[b]{0.50\linewidth}
   \includegraphics[width=\linewidth, valign=t]{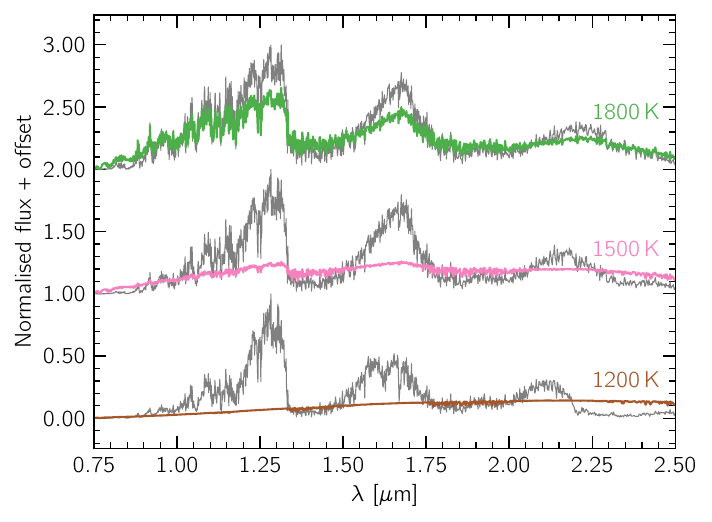}
\end{minipage}
\begin{minipage}[b]{0.50\linewidth}
    \includegraphics[width=\linewidth, valign=t]{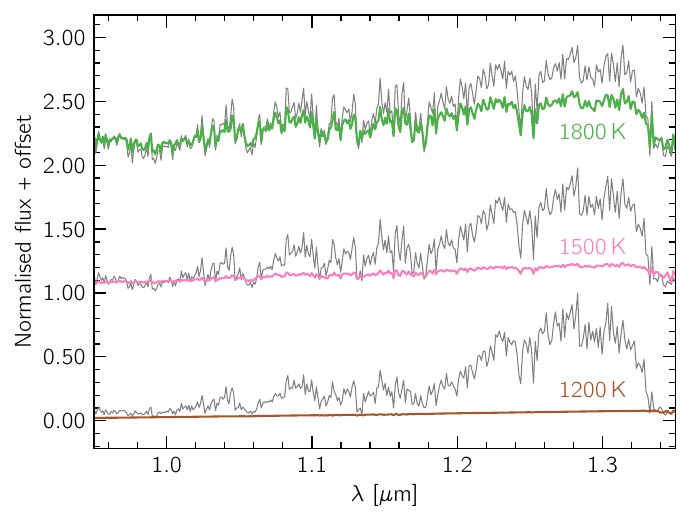}
\end{minipage}
\begin{minipage}[b]{0.50\linewidth}
    \includegraphics[width=\linewidth, valign=t]{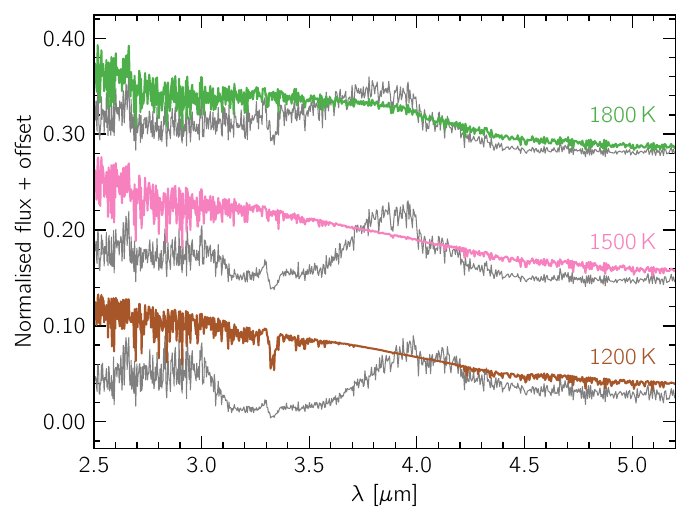}  
\end{minipage}
\begin{minipage}[b]{0.50\linewidth}    
    \includegraphics[width=\linewidth, valign=t]{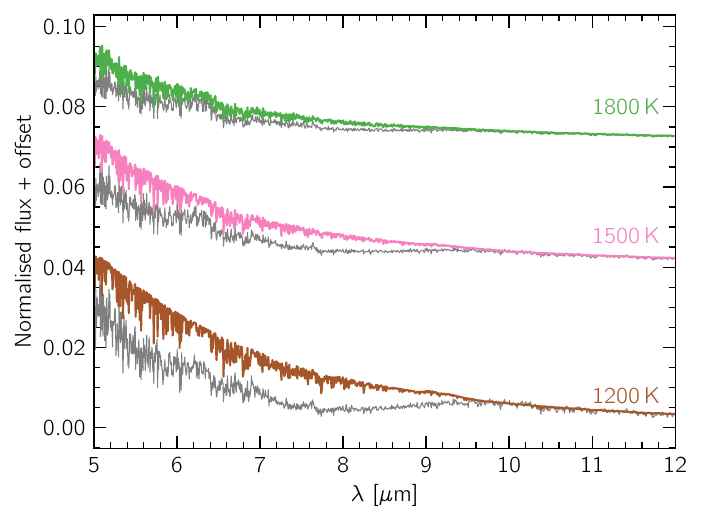}
\end{minipage}
     \caption{Synthetic spectra of \texttt{MSG} models with SiO nucleation at $T_{\rm{eff}}$$\,=\,$1800\,K (green) and 1500\,K (pink) and 1200\,K (brown), with $\log(g)=4.0$, and the respective cloud-free spectra at the same effective temperatures and $\log(g)$ in grey. The emergent fluxes are normalised with respect to the cloud-free \texttt{MSG} spectra as detailed in the caption of Fig. \ref{fig:spectra_TiO2}, and an arbitrary offset is added for clarity. In the top-left plots we show the NIR range, in the top-right the Y and J bands, in the bottom-left the MIR and the bottom-right the thermal infrared. For a reference on the location of important absorbers see Fig.~\ref{fig:spectra_TiO2}.}
     \label{fig:spectra_SiO}
\end{figure*}

Fig.~\ref{fig:pt_SiO} (left) shows a comparison between the $P_{\mathrm{gas}}-T_{\mathrm{gas}}$ profile of a \texttt{MSG} cloudy model with \ce{TiO2} nucleation (blue dashed curve) and one with SiO nucleation (orange solid curve), both at $T_{\rm{eff}}$$=$1500\,K and $\log\,(g)=4.0$. Similarly to the model with TiO$_2$ nucleation, there is a detached convective zone, which has its origin in the cloud's backwarming effect. Fig.~\ref{fig:pt_SiO} (middle) shows the average cloud particle size (solid curves) and the cloud particle number density (dashed curves) for the models shown in Fig.~\ref{fig:pt_SiO} (left). For the $P_{\mathrm{gas}}-T_{\mathrm{gas}}$ structures and cloud properties of models with SiO nucleation at other $T_{\rm{eff}}$'s, see Appendix~\ref{ap:SiO}.

If only nucleation processes are taken into account, SiO nucleation is more efficient than TiO$_2$ nucleation \citep{Lee2015}, and would therefore form more CCN overall. However, as discussed by \citet{Lee2015}, we must consider other cloud formation processes, namely growth. This is because the elements Si and O are part of many silicate materials (e.g., \ce{SiO2}[s], \ce{MgSiO3}[s], \ce{Mg2SiO4}[s]) that are already thermally stable and therefore grow efficiently as soon as the CCN form from the gas-phase. At pressures between $\sim\,10^{-7}$ to $10^{-4}$\,bar, the \ce{SiO} nucleation is more efficient than the \ce{TiO2} nucleation (see Fig.~\ref{fig:pt_SiO}, right). However, as soon as other Si-bearing cloud species start growing, the growth processes dominate over the nucleation, and the nucleation rate drops drastically. On the other hand, when TiO$_2$ nucleation is considered, the nucleation process is fairly efficient, and the growth of the Ti-bearing cloud species never dominates over nucleation. The \ce{SiO} nucleation model shows a larger cloud particle number density and smaller average cloud particle size than the \ce{TiO2} nucleation model for pressures less than $10^{-3}$\,bar where the \ce{SiO} nucleation rate is higher than in the \ce{TiO2} case (see Fig.~\ref{fig:pt_SiO}, middle). Once the \ce{SiO} nucleation rate drops, at pressures higher than $10^{-3.5}$\,bar, the cloud particle number density becomes approximately constant as no new particles are created. In comparison, at $10^{-3}$\,bar the cloud particle number density for the \ce{TiO2} nucleation case becomes larger than the \ce{SiO} case as the nucleation persists to deeper in the atmosphere. Consequently, the average cloud particle size increases in the \ce{SiO} model compared to the \ce{TiO2} nucleation model at these pressures.\footnote{We note the average cloud particle sizes obtained here are not comparable to those in \citet{Lee2015} as \texttt{MSG} is fully self-consistent, while \citet{Lee2015} do a post-processing.} This follows the principle of mass conservation: although there is less surface available for growth, the same amount of material condenses, and therefore, it must continue growing on top of the available surface, leading to larger cloud particles and smaller cloud particle number densities.

Fig.~\ref{fig:cloud_comp_SiO} shows the cloud composition and the relative gas-phase element depletion along the atmosphere for the model with SiO nucleation at $T_{\rm{eff}}$$\,=\,$1500\,K and $\log\,(g)=4.0$. The TOA has the largest difference in cloud composition between the TiO$_2$ nucleation models and the SiO nucleation models. In the TiO$_2$ models, right after TiO$_2$ nucleates at the TOA, other cloud species start growing and quickly the silicates dominate the cloud volume fraction from the very top down to approximately 0.05\,bar where Fe$_2$O$_3$[s] and Al$_2$O$_3$[s] become the dominant cloud species (Fig.~\ref{fig:cloud_comp_TiO2} middle). In the SiO models, at the TOA, SiO[s] and MgO[s] are the dominant cloud species down until the growth of magnesium-silicates and SiO$_2$ starts dominating over nucleation at $\sim 10^{-4}$ bar. Between $\sim 10^{-4}$ bar and 0.1 bar, the silicates dominate the cloud volume fraction, down until Fe[s] and Al$_2$O$_3$[s] become the dominant species at the very cloud bottom (Fig.~\ref{fig:cloud_comp_SiO} left).

Fig.~\ref{fig:spectra_SiO} shows \texttt{MSG} cloudy and cloud-free spectra at different wavelength ranges, for the models with SiO nucleation, at $T_{\rm{eff}}$ $\,=\,$ 1800\,K, 1500\,K and 1200\,K ($P_{\mathrm{gas}}-T_{\mathrm{gas}}$ profiles shown in Fig.~\ref{fig:pt_profiles}). The spectra show visible differences from the TiO$_2$ nucleation model spectra (shown in Fig.~\ref{fig:spectra_TiO2}) in all the wavelength regimes investigated. The SiO nucleation spectra are less red in the NIR (Fig.~\ref{fig:spectra_SiO} top row). For example, although small at 1500\,K, the absorption features of K and H$_2$O are still noticeable. In the MIR (Fig.~\ref{fig:spectra_SiO} bottom left), the emergent flux in the cloud-free spectra is stronger at around 4.0\,$\mu$m than the emergent flux of the cloudy spectra. 

As mentioned before, the SiO nucleation models have a larger average cloud particle size than the TiO$_2$ nucleation models. We expect the emission cross-section for magnesium-silicates to be larger for larger dust grain sizes \citep[e.g.][]{Min2004}. This, combined with the reduced cloud particle number density, results in a different cloud opacity, which gives rise to the observed spectral changes, and different $P_{\mathrm{gas}}-T_{\mathrm{gas}}$ structures (see Fig.~\ref{fig:pt_SiO}, left). Fig.~\ref{fig:cloud_comp_SiO} left panel shows the optical depth of the cloud and gas along the atmosphere at the wavelengths of 1.1\,$\mu$m and 10.0\,$\mu$m, for the model at 1500\,K. At both wavelengths, the cloud becomes optically thick around 0.1 bar (Fig.~\ref{fig:cloud_comp_SiO}, left). This is different to the \ce{TiO2} nucleation model (Fig.~\ref{fig:cloud_comp_TiO2}, left), where the cloud became optically thick around 0.4\,bar at 1.1\,$\mu$m, while it was never optically thick at 10.0\,$\mu$m. This can partially explain the spectral differences observed between the two models as we are observing different parts of the atmosphere.

\begin{figure*}[t]
\includegraphics[width=0.50\linewidth]{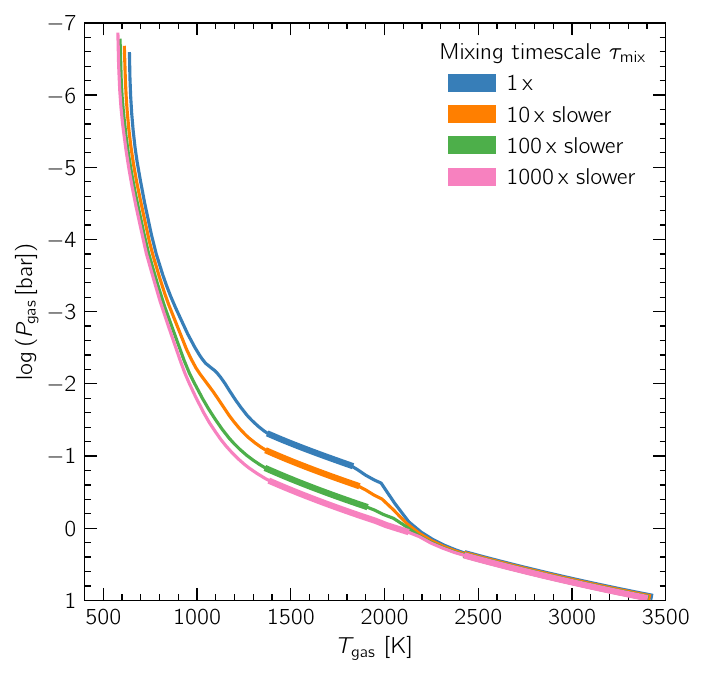}
\includegraphics[width=0.48\linewidth]{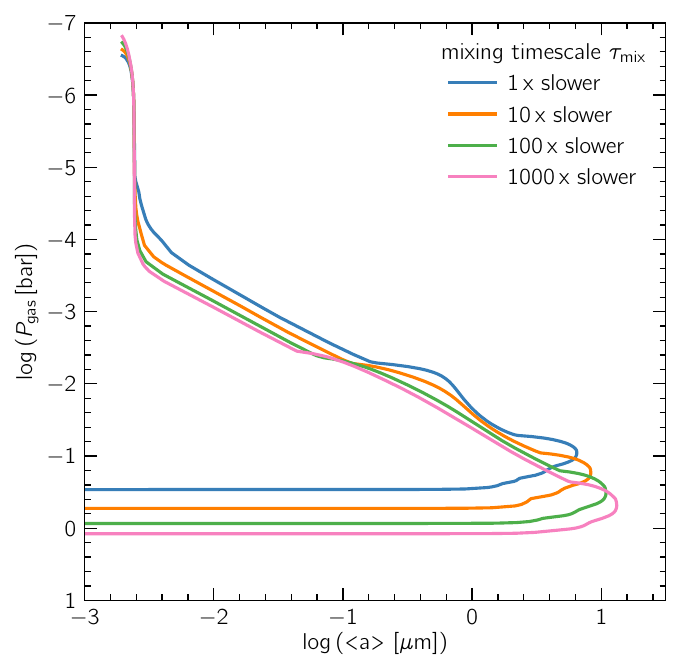}
\includegraphics[width=0.48\linewidth]{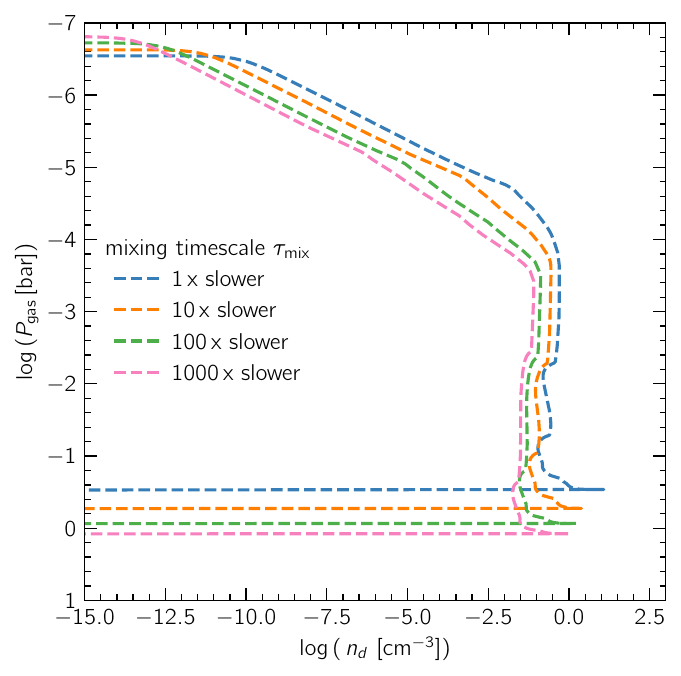}
\hspace{20pt}
\includegraphics[width=0.48\linewidth]{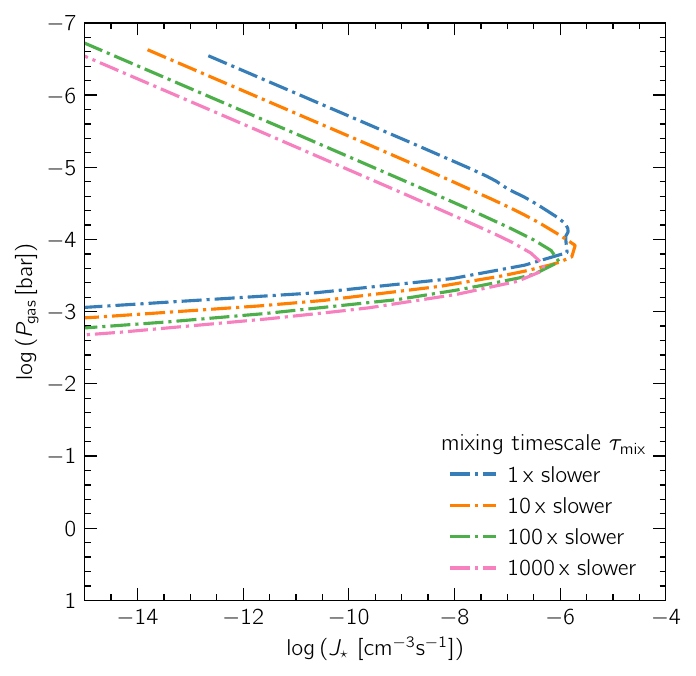}
\caption{Pressure-temperature profiles (top left) for \texttt{MSG} models with SiO nucleation, at $T_{\rm{eff}}$$\,=\,$1500\,K and $\log(g)=4.0$, and different mixing timescale scalings (1x, 10x slower, 100x slower, 1000x slower). Convective zones are plotted with thicker lines while radiative zones are plotted with thinner lines. The average cloud particle size $\langle a \rangle$ (top right), the cloud particle number density $n_d$ (bottom left) and the nucleation rate $J_\star$ (bottom right) along the atmosphere for models with SiO nucleation at $T_{\rm{eff}}$ $=$ 1500\,K and $\log(g)=4.0$, and different mixing timescale scalings.}
\label{fig:SiO_mix}
\end{figure*}
\subsection{Mixing: the effect of decreasing the mixing efficiency}
\label{sec:mixing}

We explore our assumption and parameterisation of the mixing timescale (see section~\ref{sec:SW}, equations~\ref{eq:tau_mix} and~\ref{eq:tau_mix_overshoot}) by scaling the mixing timescale up, this is reducing the efficiency of the mixing. This means the replenishment of the upper atmosphere with the cloud-forming elements happens on a slower timescale. This parameterisation breaks the self-consistency, however we find this to be the most reasonable manner to test our assumptions and see how a less efficient mixing would affect the atmosphere. Here we present the results for the models with \ce{SiO} nucleation, however we note that the models with \ce{TiO2} nucleation show the same trends and can be found in Appendix~\ref{ap:TiO2_mix}. 

\begin{figure*}[t]
\centering
\includegraphics[width=0.48\linewidth]{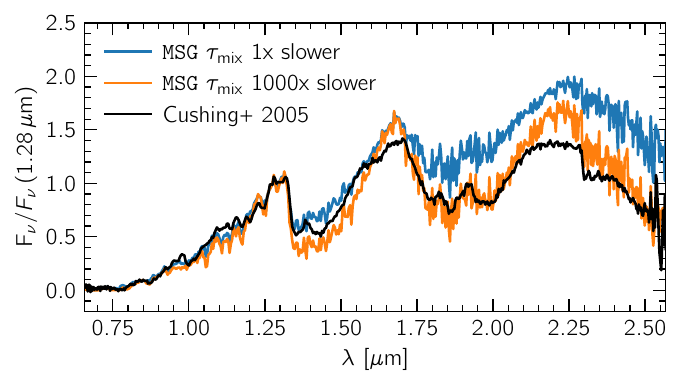}
\includegraphics[width=0.48\linewidth]{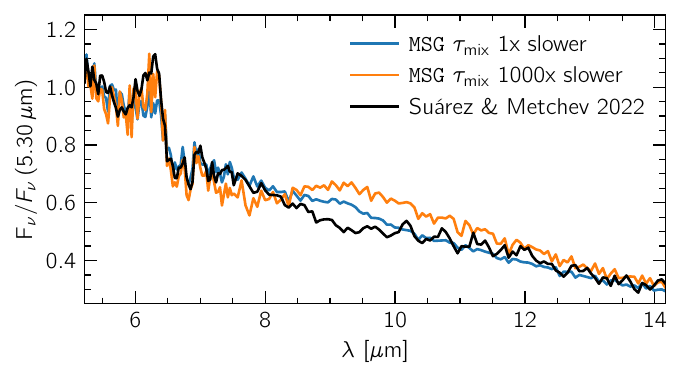}
\caption{Comparison between observed spectra of 2MASS 1507-1627 (black curves) \citep{Cushing2005, Suarez2022} and two \texttt{MSG} cloudy models with \ce{SiO} nucleation: one with fully self-consistent mixing (blue curves), and the other where the mixing timescale was scaled up by 1000 times (orange curves). The \texttt{MSG} models are at \Teff=1600\,K and log(g)=4.0.}
\label{fig:spectra_SiO_mix}
\end{figure*}
Fig.~\ref{fig:SiO_mix} (top left) shows the $P_{\mathrm{gas}}-T_{\mathrm{gas}}$ structures for the models with the less efficient mixing. Fig.~\ref{fig:SiO_mix} shows the average cloud particle size (top right), the cloud particle number density (bottom left) and the nucleation rate (bottom right) with the different mixing efficiency scalings, at $T_{\rm{eff}}$$\,=\,$1500\,K and $\log(g)=4.0$. The nucleation rate was lower in the \ce{SiO} nucleation models than in the \ce{TiO2} models. Scaling the mixing efficiency down makes the nucleation less efficient, resulting in similar behaviour to the one seen in the \ce{SiO} nucleation models compared to the \ce{TiO2} nucleation ones (section~\ref{sec:nucleation}). At the TOA, the average cloud particle size is similar in all models with different mixing efficiencies. However, the nucleation rate is slower for the decreased mixing efficiencies, and therefore, the cloud particle number density is lower. Once growth processes start dominating, i.e. when the nucleation rate drops, the average cloud particle size increases. As there are fewer CCN available for the reduced mixing efficiency cases, the cloud particles grow to slightly larger sizes. For more efficient mixing, the point of stronger growth onset is higher in the atmosphere than for the less efficient mixing cases (see Fig.~\ref{fig:SiO_mix}, top right). As previously shown by \citet{Samra2023_WASP96b}, more efficient mixing supports a higher cloud deck. The location of the cloud deck is also a consequence of the $P_{\mathrm{gas}}-T_{\mathrm{gas}}$ structure (see Fig.~\ref{fig:SiO_mix}, top left), which is hotter for more efficient mixing due to the different cloud and gas opacities (with the differences being caused by the change in average cloud particle size, cloud composition and gas-phase element abundances). For detailed figures of the average cloud composition and gas-phase element abundances see Appendix~\ref{ap:SiO_mix}.

The effect of the scaling on the $P_{\mathrm{gas}}-T_{\mathrm{gas}}$ structure is similar to that of SiO nucleation models compared to the \ce{TiO2} nucleation models (see Fig.~\ref{fig:pt_SiO} left). Between approximately 0.001\,bar and 1\,bar, the models with less efficient mixing are cooler overall. Fig.~\ref{fig:spectra_SiO_mix} shows synthetic spectra in the NIR and MIR for the model resulting from scaling the mixing timescale up by 1000 times (orange), compared to the spectrum with no scaling (blue), and observed spectra of 2MASS 1507-1627. In the NIR, the model with the reduced mixing efficiency better matches the observed data compared to the fully self-consistent model. However, in the MIR, particularly for where we would expect the silicate feature, the model with reduced mixing efficiency is an even worse fit than the fully self-consistent model. This is not surprising as the average cloud particle size in the observable atmosphere increases with the reduced mixing efficiency (Fig~\ref{fig:SiO_mix}, top left). Larger cloud particle sizes are known to have a less prominent, or even no silicate feature \citep[e.g.][]{Min2004, Luna2021}. The NIR spectra is likely less red due to a decrease in the cloud particles number density, a result of both the reduced mixing efficiency and the subsequently  reduced nucleation rate.

\subsection{The optical properties of the cloud}
\label{sec:cloud_opt_properties} 
\begin{figure}
    \centering
    \includegraphics[width=\linewidth]{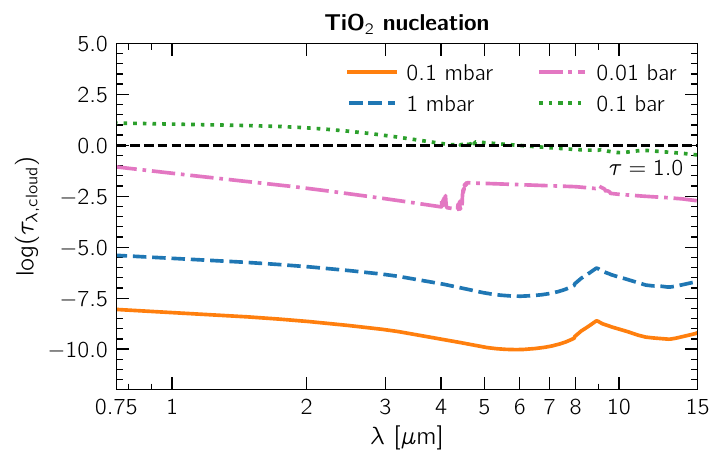}
    \includegraphics[width=\linewidth]{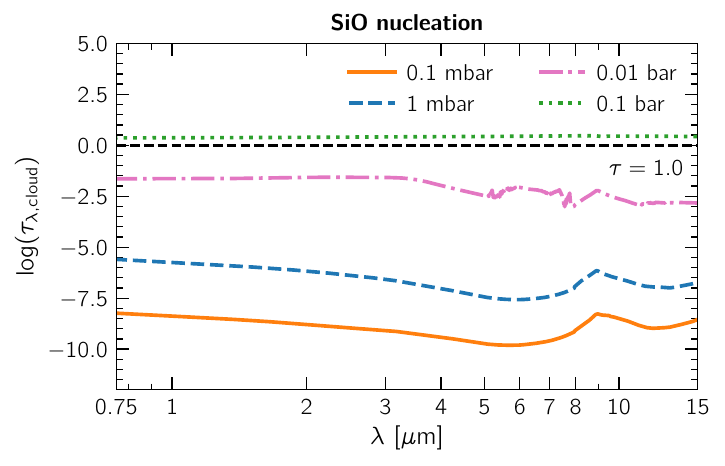}
    \caption{The cloud optical depth as a function of wavelength, at four different heights (pressures) in the atmosphere, for two \texttt{MSG} models at \Teff = 1500\,K, log(g) = 4.0, with \ce{TiO2} nucleation (top) and SiO nucleation (bottom). The $\tau = 1.0$ level is plotted as the black dashed curve for reference.}
    \label{fig:tau_vs_lambda}
\end{figure}

To investigate why we miss the silicate feature in our model spectra, we explore some optical properties of the cloud, namely the optical depth and the single scattering albedo. Fig.~\ref{fig:tau_vs_lambda} shows the optical depth of the cloud as a function of wavelength, at different pressures in the atmosphere, for two \texttt{MSG} models at \Teff = 1500\,K, log(g) = 4.0.  One model assumes  \ce{TiO2} nucleation, while the other assumes SiO nucleation (see Fig.\,\ref{fig:pt_SiO} for the corresponding cloud properties, and Figs.\,\ref{fig:cloud_comp_TiO2} and \ref{fig:cloud_comp_SiO} (middle panel) for the cloud compositions). For both models, the cloud is optically thin at the TOA, down to about 0.1\,bar where it becomes partially/totally (\ce{TiO2}/SiO) optically thick in the wavelength regime explored. In both cases, it is possible to see the silicate feature at 8-10\,$\mu$m at pressures equal and smaller than 1\,mbar. At these pressures, the average cloud particle size is 0.01-0.1\,$\mu$m for which a silicate feature is expected. For the SiO nucleation model, there is still a visible silicate feature at 0.01\,bar, however this is not the case for the \ce{TiO2} model. When the cloud becomes optically thick in both models, the silicate feature is no longer visible.

\begin{figure}
    \centering
    \includegraphics[width=0.98\linewidth]{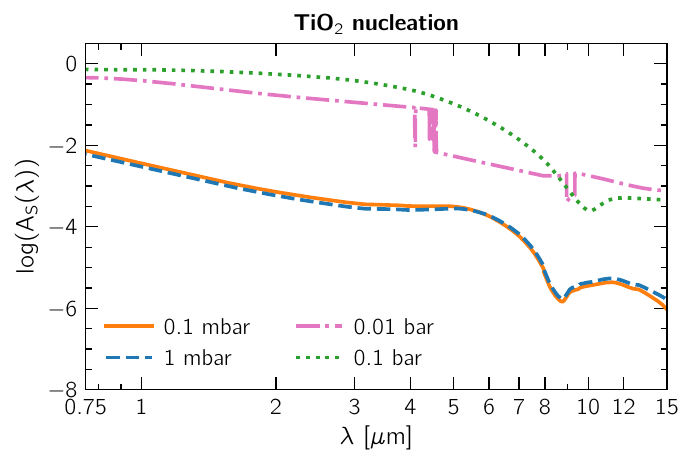}
    \includegraphics[width=0.98\linewidth]{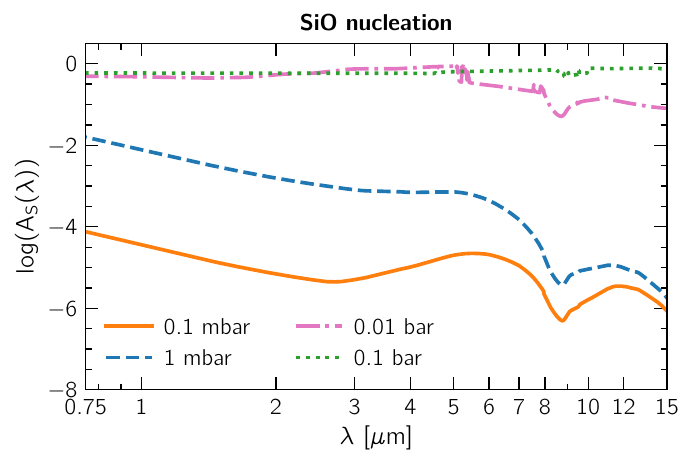}
    \caption{The single scattering albedo of the cloud particles as a function of wavelength, at four different heights (pressures) in the atmosphere, for two \texttt{MSG} models at \Teff = 1500\,K, log(g) = 4.0, with \ce{TiO2} nucleation (top) and SiO nucleation (bottom). The single scattering albedo is the ratio of the scattering efficiency to the extinction efficiency.}
    \label{fig:ssa}
\end{figure}

Fig.~\ref{fig:ssa} shows the cloud particles' single scattering albedo (A$_{\mathrm{S}}$) as a function of wavelength, at different pressures in the atmosphere, for the same two models presented in Fig.~\ref{fig:tau_vs_lambda}. For both models the silicate feature is clearly visible, between 8-9\,$\mu$m, for 0.1-1\,mbar. For the \ce{TiO2} model, A$_{\mathrm{S}}$ does not vary notably for 0.1-1\,mbar because the cloud composition and the average cloud particle size do not change significantly (see Fig.~\ref{fig:cloud_comp_TiO2} middle, Fig.~\ref{fig:cloud_size_nd_TiO2} left). For the SiO model, between 0.1 and 1\,mbar, A$_{\mathrm{S}}$ increases up to about 2 orders of magnitude in the NIR. This is likely due to compositional changes of the cloud particles, as well as an increase in the average cloud particle size of about 1 order of magnitude (Fig.~\ref{fig:pt_SiO}, middle). Deeper in the atmosphere, at 0.01-0.1\,bar, A$_{\mathrm{S}}$ increases significantly for both models in the NIR, due to the increase in the average cloud particle size. For the SiO nucleation model, this increase also occurs in the MIR. This is because the average cloud particle sizes become comparable to the radiation's wavelength, entering the Mie scattering regime. A notable difference between the two models is the MIR regime for pressures above 0.01\,bar. This arises from differences in both the average cloud particle sizes, which are larger for the SiO model at this part of the atmosphere, as well as compositional differences for the cloud particles. In particular, it is possible to see a feature just before 10\,$\mu$m at 0.1\,bar for the \ce{TiO2} nucleation model, which in this case is not due to silicate absorption but rather \ce{Al2O3}[s] scattering (see cloud composition in Fig.~\ref{fig:cloud_comp_SiO} middle; see Figure~5 of \citet{Luna2021}). 

Although the trace of the silicate cloud is seen when looking only at the cloud optical properties, this is not seen in the synthetic spectra as these features occur at altitudes where the cloud is not optically thick, and where the gas is generally optically thicker than the cloud. Our self-consistent model does not produce enough small cloud particles at the correct altitudes for the silicate feature to be visible. In section~\ref{sec:silicate} we discuss what could be affecting this in our modelling framework.

\section{Discussion}
\label{sec:discussion}
\subsection{Convergence challenges of self-consistent brown dwarf models}
\label{sec:convergence}
Besides the convergence challenges we face when considering the cloud opacity in a self-consistent scheme (see section~\ref{sec:control}), there are a number of other convergence challenges in our models.

In cloud-free \texttt{MSG} models, we face challenges with convergence at effective temperatures between 1600\,K and 1200\,K due to so called ``opacity cliffs'' \citep{Mukherjee2023}. ``Opacity cliffs'' are regions where the gas opacity changes quickly with small changes in temperature or pressure. These cliffs are also seen in Rosseland mean opacities (Figure 2 in \citet{Freedman2008}; Figure 3 in \citet{Freedman2014}). We note \texttt{MARCS} models are computed over a Rosseland optical depth scale. The cloud-free models at effective temperatures between  1600\,K and 1200\,K often end up oscillating between two temperature corrections of the same amplitude but with different signs. For these cloud-free cases, a given model is more likely to reach convergence if the applied temperature correction is both smaller than the current temperature oscillation, and larger than the convergence criteria applied. This works by introducing a temperature correction which breaks out of the oscillation. However, this does not seem to help in the \texttt{MSG} cloudy models case. Besides an ``opacity cliff'' due to the gas opacity, the cloud opacity also contributes to the cliff as it changes rapidly with pressure, especially at lower effective temperatures. 

In addition, we face a challenge when computing the cloud opacities as the vertical scale in \texttt{MARCS} is different from that of \texttt{DRIFT}. Unless the input values are exactly the same, \texttt{DRIFT} never outputs the same pressure points because it computes the necessary atmospheric height steps within its numerical methods, meaning the number of atmospheric layers out of a \texttt{DRIFT} run is variable. On the other hand, the number of atmospheric layers in \texttt{MARCS} is fixed and defined on a Rosseland optical depth scale. This introduces non-ideal interpolation errors when interpolating the necessary \texttt{DRIFT} outputs to perform the Mie theory and EMT calculations. The major problem is rooted in the fact that due to the ``opacity cliffs'' mentioned above, a small change in pressure leads to a large change in the Rosseland optical depth. This is exacerbated at lower effective temperatures. Following this, the chosen points for the interpolation are almost always different in consecutive iterations. 
Although a Rosseland optical depth scale works well for hotter objects, it is not the best choice for cooler objects where the cloud opacity has a large influence. In the future, it might be beneficial to switch this treatment in \texttt{MSG} to use a scale in gas pressure, or use an already existent atmospheric model with the scale in gas pressure.

\subsection{Silicate cloud feature}
\label{sec:silicate}
If silicate clouds are present in brown dwarf atmospheres, they are expected to show a significant absorption feature at about 10$\,\mu$m. \citet{Suarez2022} presented an analysis of \textit{Spitzer} MIR spectra of 113 field M5-T9 dwarfs. They find silicate absorption starts to appear at the L2 spectral type, is strongest in L4-L6 dwarfs, and disappears past L8 dwarfs. Nevertheless, the silicate absorption feature is not ubiquitous and can be missing at any L sub-type. More recently, a silicate absorption feature was detected in the planetary mass companion VHS\,1256\,b with JWST MIRI-MRS \citep{Miles2023, Petrus2024}. 

In the models presented here, we do not see a silicate absorption feature at 10$\,\mu$m. In section~\ref{sec:cloud_opt_properties}, we look at the cloud's optical depth and single scattering albedo. As previously mentioned, when exploring these optical properties it is possible to see the existence of the silicate feature at pressures around 0.1-1\,mbar, both in the optical depth and the single scattering albedo. However, this feature is not visible in the synthetic spectra because the gas tends to be optically thicker than the cloud at the relevant wavelengths and altitudes. In addition to this, at the altitudes where the cloud becomes optically thick around the 10\,$\mu$m feature, the cloud particles are no longer small enough to give rise to a silicate feature. For example \citet{Min2004}, and more recently \citet{Luna2021}, show how smaller cloud particles ($\sim 0.1-1.0 \, \mu m$) have a more outstanding silicate absorption feature around $10\,\mu$m.  Our results indicate that the models do not have enough small ($\sim 0.1-1.0 \, \mu m$) cloud particles at lower pressures, for a silicate feature to be observable. We speculate two reasons why this could be the case:

\begin{enumerate}
\item The nucleation rate is underestimated. If this is the case, this implies we are not producing sufficient CCN at the TOA for the cloud particles to grow on. If the nucleation rate were to be increased, we would expect an increase in available CCN, meaning an increase in the cloud particle number density, and due to mass conservation a decrease in the average cloud particle size. This would imply an increase in smaller cloud particles at lower pressures. Unfortunately due to the self-consistent nature of the cloud formation model, it is not possible to scale the nucleation rate up and down (like it is possible for the mixing timescale) to test this case. We tested reducing the number of monomers in the CCN, $N_\ell$ (see equation~\ref{eq:elem_cons}), from 1000 to 10, however this did not make a visible difference in the silicate feature, although it produced slightly less red spectra overall.
\item The mixing efficiency is underestimated. If this is the case, the TOA is not being replenished with non-depleted gas abundances fast enough to support efficient nucleation. However, increasing the amount of elements available at the TOA will not only affect the nucleation rate but also the growth rate of the cloud particles, as there will be more material available for growth. The number density of cloud particles will increase, however without appropriate testing it is not possible to predict what the average cloud particle sizes would be. Such a scenario should be tested in future modelling. 

\end{enumerate}

We note we find a better match with observations in the NIR when reducing the mixing efficiency. However, less efficient mixing made the fit in the MIR worse. This could be an indication that we are observing patchy clouds, and a single 1D model is not enough to explain the observed spectra. This idea matches the with the many spectroscopic and photometric variability observations in brown dwarfs \citep[e.g.][]{Biller2024, Vos2022, Vos2019, Apai2013}. \citet{Burningham2021} and \citet{Vos2023} find that in their retrievals on archival data there is a preference for patchy clouds. 

Using the microphysical cloud model \texttt{CARMA}, which is based on a bin-scheme, \citet{Powell2018} find the particle size distributions in hot Jupiters are often bimodal. In the future, different particle size distribution functions should be considered in the \texttt{MSG} framework when computing the cloud opacity to evaluate if the silicate feature can be replicated in this manner, and what is the impact of having a particle size distribution in synthetic spectra. 

\subsection{Detached convective zones, the L-T dwarf transition and brown dwarf spectroscopic variability}
\label{sec:conv}
In both models with \ce{TiO2} nucleation and \ce{SiO} nucleation, detached convective zones appear for effective temperatures below 1600\,K due to the cloud's backwarming effect. This is also observed in other model grids which consider cloud formation \citep[e.g.][]{Saumon2008, Malik2019, Morley2024}{}{}. In this work, as described at the end of section~\ref{sec:SW}, we set the mixing timescale $\tau_{\rm{mix}}$ at the detached convective layer to the value of $\tau_{\rm{mix}}$ at the top of the radiative zone just below. This assumption misses two important considerations: (1) we fail to consider the full motion of the gas due to convection, which can accelerate the element replenishment at the TOA; (2) we do not consider the cloud particles will be dragged by the moving gas elements. 

\citet{Witte2011} modified the dust moment equations in the \texttt{DRIFT-PHOENIX} models to consider the effect of the convective motion in the cloud particles. They find the gas velocities can exceed the cloud particle settling velocities by several orders of magnitude. This dynamic interaction disrupts cloud layers at the base by splitting the cloud particles: approximately half are driven into the upper atmosphere, where they continue to grow and eventually settle back to their initial altitude, while the other half are pushed deeper into the atmosphere, where they evaporate. This convective cycling reduces the number of cloud particles locally, stopping a trend of reddening in the NIR. However, from effective temperatures below 1400\,K, they find strong numerical oscillations in their model and struggle to find convergence.

This type of cloud disruption has been hypothesised as a potential mechanism driving the L-T dwarf transition at about effective temperatures of $\sim$1400 K, where brown dwarfs shift rapidly from red to blue NIR colours, and a significant brightening is observed in the J band of early T dwarfs \citep[e.g.][]{Allard2001, Marley2002, Tsuji2002, Kirkpatrick2005}. \citet{Burgasser02} and \citet{Burrows06} discuss convection as a potential disruption mechanism of clouds in brown dwarfs, and as the possible mechanism that gives rise to the L-T transition in more detail. 

In addition to the L-T dwarf transition, large observational surveys have found high-amplitude spectroscopic variability to be ubiquitous across the entire L-T spectral sequence \citep{Metchev2015, Vos2019, Vos2022, Liu2024}. Clouds are also thought to be the primary cause of these observed variabilities. Using a simple time-dependent, self-consistent, 1D model, \citet{Tan2019} find that spontaneous atmospheric variability can be driven by radiative cloud feedback, with cycles of cloud growth and dissipation tied to detached convective zones.

Alternative mechanisms for the origin of the L-T transition include chemical instabilities, such as diabatic convection from carbon chemistry \citep[e.g.][]{Trembling2016, Tremblin2019}, which may act alongside or independently of cloud processes. Current JWST observations aim to disentangle these effects by studying variable brown dwarfs \citep[e.g. PI: Biller, B. Program ID: 2965;][]{Biller2023_jwst}.

In future work, we find it crucial to include the convective motion consideration in the dust moments equation in order to investigate if it is a plausible mechanism for the origin of the L-T transition in our model framework. It is necessary to understand if our control factor treatment of the cloud opacity (section~\ref{sec:control}) would give rise to the same numerical oscillations found by \citet{Witte2011}, or if there is a static solution, and if we are able to replicate the bluer NIR colours of early T-dwarfs. Although challenging, exploring the combined effect of chemical instabilities and microphysical cloud formation would be interesting.

\subsection{Nucleation}
\label{sec:nuc_disc}

Nucleation is the initial stage of cloud formation in gas atmospheres, involving the formation of molecular clusters through gas-gas reactions \citep{Gail1984,Helling2013}. Identifying the species that nucleate efficiently in astrophysical environments depends on the balance between the abundance of constituent elements and the binding energies of the clusters \citep{Helling2013}. Various potential nucleation species in substellar environments have been proposed. For warmer environments, candidates include \ce{TiO2} \citep{Jeong2000, Sindel2022}, \ce{SiO} \citep{Gail2013, Bromley2016}, \ce{Al2O3} \citep{Lam2015, Gobrecht2022}, and vanadium oxides like \ce{VO2} and \ce{V2O5} \citep{Lecoq-Molinos2024}. In cooler atmospheres, salts (e.g., NaCl, KCl) and metal sulfides (e.g., ZnS) are potential candidates \citep{Lee2018, GaoBenneke2018}. Following prior studies on hot Jupiters \citep[e.g.,][]{Helling2006, Helling2019_HATP7b}, we adopt \ce{TiO2} and \ce{SiO} as nucleation species in our models, consistent with the \texttt{DRIFT-PHOENIX} framework \citep{Helling2008_consistent}.

In this work, we test two nucleation species individually, \ce{TiO2} and \ce{SiO}, to explore the impact of different nucleation rates. It is expected that multiple nucleation species will form CCN simultaneously in substellar atmospheres \citep[e.g.]{Lee2018, Helling2023_gridpaper}, however, as stated previously the efficiency of nucleation species can vary significantly. Previous works applying \texttt{DRIFT} \citep[e.g.][]{Lee2015,Helling2021_WASP_Comparison} have shown that \ce{SiO} nucleates more efficiently than \ce{TiO2} in the regions where they can both nucleate. Therefore, we use \ce{SiO} to test the more efficient nucleation case, and \ce{TiO2} to test the less efficient nucleation case.

Nucleation modelling remains uncertain due to sensitivities to theoretical approaches and a lack of laboratory data. Although we tested both \ce{TiO2} and \ce{SiO} as nucleation species, we cannot make any conclusions regarding which species is more likely to nucleate. The major reasoning for testing one species against the other 
Comparisons of classical and non-classical nucleation theories show significant differences in predicted nucleation rates, varying by two orders of magnitude for \ce{TiO2} \citep{Sindel2022} and 15 orders for \ce{V2O5} \citep{Lecoq-Molinos2024}. These studies highlight the need for accurate surface energy measurements at relevant temperatures. Observationally, nucleation has not been detected in substellar atmospheres, but JWST/MIRI-LRS may provide evidence by detecting molecular clusters like (Al$_2$O$_3$)$_N$ and (TiO$_2$)$_N$. A JWST Cycle 3 observation targeting these features on WASP-76\,b is scheduled \citep{Baeyens2024_JWST_Prop}.

\subsection{Mixing}
\label{sec:mix_disc}
In our mixing prescription, we assume there is element replenishment at the TOA from a reservoir in the deep atmosphere. We assume the elements are transported from this reservoir by convective mixing (see section~\ref{sec:SW}). However, we do not consider that these elements could take part in other cloud formation processes as they travel up the atmosphere (e.g. nucleation or growth). We could be overestimating the true mixing efficiency by not considering the possibility of these other processes happening. This is perhaps why our models with a scaled-up mixing timescale, and therefore less mixing efficiency, tend to improve the agreement with observations (Fig.~\ref{fig:spectra_SiO_mix}).

Scaling of the mixing timescale in \texttt{DRIFT} has previously been applied in an attempt to improve the agreement with observations. Pre-JWST observations of the warm Saturn exoplanet WASP-96\,b with VLT/FORS2 \citep{Nikolov2018} revealed the Na I line at $\sim0.6\,{\rm \mu m}$ with broad wings leading the authors to conclude they observe a cloud-free atmosphere. WASP-96\,b is a JWST ERO target and was observed with JWST NIRISS/SOSS \citep{Pontoppidan2022, Radica2023, Taylor2023}. \citet{Radica2023} and \citet{Taylor2023} find they do not require a grey cloud deck in their retrievals to match the observations. However, they have to include a Rayleigh scattering slope. \citet{Samra2023_WASP96b} predict condensate clouds in the atmosphere of WASP-96\,b and find their \texttt{DRIFT} models predict a cloud top, which is inconsistent with the broadened Na I line seen in the VLT/FORS2 observations. To address this, \citet{Samra2023_WASP96b} scaled the mixing timescale, finding that increasing the mixing timescale by a factor of 100$\times$ shifts the cloud top to higher pressures of $0.01$ bar, which is more consistent with observations.

Several other cloud models \citep[e.g.][]{Ackerman2001,Gao2018, Ormel2019}{}{} assume the vertical mixing to be diffusive. This diffusive mixing is parameterised by the eddy diffusion coefficient $K_{zz}$ [cm$^2$/s], a coefficient which approximately encompasses a number of large-scale transport processes in substellar atmospheres, such as convection and atmospheric circulation. The estimation of $K_{zz}$ often depends on the type of object being modelled. For example, in \citet{Morley2024} \citep[following][]{Ackerman2001}, $K_{zz}$ is calculated using mixing length theory while also considering the energy transported by radiation. Overall, $K_{zz}$ is often used to describe different types of transport, and it can be difficult to pinpoint the physics that it truly probes.  

\citet{Mukherjee2024} find they require a low $K_{zz}$ in the convective region of their \texttt{Sonora Elf Owl} models to fit observed spectra of T dwarfs with effective temperatures between 500\,K-1000\,K. We note the \texttt{Sonora Elf Owl} models do not take clouds into consideration, but disequilibrium chemistry instead. They find the $K_{zz}$ computed from mixing length theory is of the order of $10^8$  cm$^2$/s, while the best fits are for values of  $K_{zz}$ between $10^1$-$10^4 \,$ cm$^2$/s \citep[see Figure 15 in][]{Mukherjee2024}.  To scale down the value of $K_{zz}$ in the convective regions, \citet{Mukherjee2024}  reduce the mixing length by a factor of 10. In the \texttt{MSG} models, we find we require less efficient mixing to better fit observed NIR spectra of L dwarfs. The mixing timescale we compute in the deepest convective regions (Equation~\ref{eq:tau_mix}) follows mixing length theory. Following equation~\ref{eq:tau_mix}, we can estimate $K_{zz}$ by considering $K_{zz} = v_{\mathrm{conv}} \, H_p = H_p^2 / \tau_{\mathrm{mix}}$.  We find values of $K_{zz}$ in the deep convective regions of the order of $10^{10}$ cm$^2$/s. The models which best fit the NIR spectra have the mixing timescale scaled up by 1000 times, meaning in these cases $K_{zz}$ is of the order of $10^7$ cm$^2$/s, a few orders of magnitude larger than the best fit  $K_{zz}$ values found by \citet{Mukherjee2024}  for T dwarfs. Retrieved values of  $K_{zz}$ for the directly imaged planet HR\,8799\,e \citep{Molliere2020} and the L dwarf  DENIS J0255-4700 \citep{deRegt2024} are comparable to those found in our work.

Overall,  it is difficult to constrain the physics of the mixing, and this will likely remain an uncertain parameter within 1D self-consistent cloudy models like the ones presented here.

%__________________________________________________________________
\section{Summary}
We have presented a new grid of \texttt{MSG} cloudy substellar model atmospheres at effective temperatures between 2500\,K and 1200\,K and at a surface gravity of $\log(g)=4.0$. We have presented a new convergence algorithm for the coupling of \texttt{MARCS} to \texttt{DRIFT} in comparison to the grid presented in \citet{Juncher2017}. The new algorithm is based on a control factor that controls the change in cloud opacity and the gas-phase element abundances to avoid unwanted numerical oscillations. We present pressure-temperature profiles, cloud properties including the cloud composition along the atmosphere and the average cloud particle sizes, and model spectra. 

Our models, which consider \ce{TiO2} nucleation, show spectra that are significantly redder in the NIR than the currently known population of substellar atmospheres. We have explored the effect of changing the CCN from \ce{TiO2} to \ce{SiO}. Changing the CCN to \ce{SiO} makes the spectra appear less red in the NIR. Additionally, we investigated the effect of making the atmospheric mixing less efficient. The models with reduced mixing also appear less red in the NIR. Overall, our models present a strong cloud continuum which does not match observations. 

We find detached convective zones in models at effective temperatures below or equal to 1600\,K. The detached convective zone originates from the backwarming effect by the cloud particles. The mixing treatment we use does not consider the effect of the convection motion on the cloud particles. We propose that it is crucial to consider the convective motion in future work. This is because it has been argued the L-T transition can have its origin in cloud clearing caused by such a convective motion \citep{Burgasser02, Burrows06}.

Our models do not present the expected silicate absorption feature in the MIR between 9 and 11\,$\mu$m. We argue this is potentially due to an underestimation of the nucleation rates and/or an underestimation of the mixing efficiency.

It is not expected that a single 1D model will fit the observed spectra following the consistent observations of spectroscopic and photometric variability in brown dwarfs. This should be taken into consideration in future works, where different 1D models are combined to test if a patchy clouds scenario can better reproduce the observations.

\begin{acknowledgements}
We thank the anonymous reviewer for their comments, which improved the manuscript. We are grateful to Peter Woitke, Paul Mollière, Dominic Samra, Antonia von Stauffenberg and Helena Lecoq-Molinos for interesting discussions. BCE, ChH, FAM and UGJ are part of the CHAMELEON MC ITN EJD which received funding from the European Union‘s Horizon 2020 research and innovation programme under the Marie Sklodowska-Curie grant agreement no. 860470. DAL acknowledges financial support from the Austrian Academy of Sciences. UGJ acknowledges funding from the Novo Nordisk Foundation Interdisciplinary Synergy Programme grant no.1716 NNF19OC0057374. The Tycho supercomputer hosted at the SCIENCE HPC center at the University of Copenhagen was used for supporting this work.
\end{acknowledgements}

\bibliographystyle{aa}
\bibliography{bib_file}
\onecolumn
\begin{appendix}
\FloatBarrier
\section{Continuum opacity sources and molecular \& atomic line lists references}
\label{ap:opac_data}
\begin{table}[hbt!]
\renewcommand{\arraystretch}{1.2}
\caption{Continuum opacity data sources.}             % title of Table
\label{tab:continuum}      % is used to refer this table in the text
\centering
                  % used for centering table
\begin{tabular}{l l l } % centered columns (4 columns)

\hline\hline          % inserts double horizontal lines
%\noalign{\smallskip}
Ion & Process & Reference \\    % table heading
\hline                        % inserts single horizontal line
%\noalign{\smallskip}
H$^{-}$ & b-f & \citet{Hmbf}  \\      % inserting body of the table
H$^{-}$ & f-f & \citet{Hmff}      \\
H\,{\tiny\uppercase\expandafter{\romannumeral 1\relax}}  & b-f, f-f & \citet{Hcont} \\
H\,{\tiny\uppercase\expandafter{\romannumeral 1\relax}}\normalsize+H\,{\tiny\uppercase\expandafter{\romannumeral 1\relax}} & CIA & \citet{HplusH} \\
H$_2^-$& f-f & \citet{H2minus}    \\
H$_2^+$& f-f & \citet{H2plus}    \\
He$^-$& f-f & \citet{Heminus}, \citet{Heminus_2}     \\
He\,{\tiny\uppercase\expandafter{\romannumeral 1\relax}}& f-f & \citet{Peach1970}    \\
C\,{\tiny\uppercase\expandafter{\romannumeral 1\relax}}& f-f & \citet{Peach1970}    \\
Mg\,{\tiny\uppercase\expandafter{\romannumeral 1\relax}}& f-f & \citet{Peach1970}    \\
Al\,{\tiny\uppercase\expandafter{\romannumeral 1\relax}}& f-f & \citet{Peach1970}    \\
Si\,{\tiny\uppercase\expandafter{\romannumeral 1\relax}}& f-f & \citet{Peach1970}    \\
e$^{-}$& scattering & \citet{Mihalas1978}    \\
H$_2$& scattering & \citet{H2scat}    \\
%\noalign{\smallskip}
\hline                                   %inserts single line
\end{tabular}
\tablefoot{Bound-free processes are denoted by b-f. Free-free processes are denoted by f-f. Collision induced absorption processes are denoted by CIA.}
\end{table}

\begin{table}[hbt!]
\renewcommand{\arraystretch}{1.2}
\caption{Molecular and atomic line lists data sources.}             % title of Table
\label{tab:mol_opac}      % is used to refer this table in the text
                       % used for centering table
\centering
\begin{tabular}{l l | l l}        % centered columns (4 columns)         % inserts double horizontal lines
\hline\hline
Molecule/atom & Reference & Molecule/atom & Reference \\    % table heading                       % inserts single horizontal line
\hline
AlCl & \citet{AlCl_Yousefi}  &  MgH & \citet{MgH_Gharib}\\
AlH & \citet{AlH_Yurchenko} & NaCl & \citet{NaCl_KCl_Barton} \\
AlO & \citet{AlO_Patrascu} & NaH & \citet{NaH_Rivlin}  \\
$\mathrm{C_2}$ & \citet{C2_Yurchenko} & NH & \citet{NH_Brooke14, NH_Brooke15, NH_Fernando} \\
CaH  & \citet{CaH_MgH_Owens} & NH$_3$ & \citet{NH3_Coles}\\
CH & \citet{CH_Masseron} & NO & \citet{NO_Hargreaves} \\
CH$_4$ & \citet{CH4_Yurchenko2017} & NS & \citet{SN_Yurchenko} \\
CN & \citet{CN_Brooke} & OH & \citet{OH_Brooke, OH_Yousefi}\\
CO & \citet{CO_Li} &  SiH & \citet{SiH_Yurchenko}  \\
CO$_2$ & \citet{CO2_Yurchenko} & SiO & \citet{SiO_Yurchenko} \\
CS & \citet{CS_Paulose} & SiS & \citet{SiS_Upadhyay}\\
FeH & \citet{FeH_Wende} & SH & \citet{SH_Gorman}\\
H$_2$CO & \citet{H2CO_AlRefaie} & SO$_2$ & \citet{SO2_Underwood} \\
H$_2$O & \citet{H2O_Polyansky} & TiH & \citet{TiH_Burrows} \\
HCN & \citet{HCN_Barber} & TiO & \citet{TiO_McKemmish} \\
KCl & \citet{NaCl_KCl_Barton} & VO & \citet{VO_McKemmish}\\
LiCl & \citet{LiCl_Bittner} & Na & \citet{Allard2019}\\
LiH & \citet{LiH_Coppola} & K & \citet{Allard2019}\\
\hline                              %inserts single line
\end{tabular}
%\tablefoot{}
\end{table}

\clearpage
\clearpage
\section{Chemical surface reactions assumed to form the cloud particles}
\label{ap:surfacereact}
\begin{table*}[hbt!]
\centering
\caption{Chemical surface reactions $r$ assumed to form the solid materials s.}
\label{tab:chemreak}
\resizebox{14.0cm}{!}{

\begin{tabular}{c|c|l|l}
{\bf Index $r$} & {\bf Solid s} & {\bf Surface reaction} & {\bf Key species} \\
\hline 
1 & TiO$_2$[s]          & TiO$_2$ 
       $\longrightarrow$ TiO$_2$[s]                  & TiO$_2$ \\ 
2 & rutile              & Ti + 2 H$_2$O 
       $\longrightarrow$ TiO$_2$[s] + 2 H$_2$        & Ti     \\
3 & (1)                 & TiO + H$_2$O  
       $\longrightarrow$ TiO$_2$[s] + H$_2$          & TiO     \\ 
4 &                     & TiS + 2 H$_2$O
       $\longrightarrow$ TiO$_2$[s] + H$_2$S + H$_2$ & TiS     \\
\hline 
5 & SiO$_2$[s]          & SiH + 2 H$_2$O
       $\longrightarrow$ SiO$_2$[s] + 2 H$_2$ + H    & SiH \\ 
6 & silica              & SiO + H$_2$O 
       $\longrightarrow$ SiO$_2$[s] + H$_2$          & SiO     \\ 
7 &  (3)                & SiS + 2 H$_2$O 
       $\longrightarrow$ SiO$_2$[s] + H$_2$S + H$_2$ & SiS     \\
\hline 
8 & SiO[s]              & SiO 
       $\longrightarrow$ SiO[s]                  & SiO \\
9 & silicon mono-oxide  & 2 SiH + 2 H$_2$O 
       $\longrightarrow$ 2 SiO[s] + 3 H$_2$      & SiH   \\
10 & (2)                & SiS + H$_2$O 
       $\longrightarrow$ SiO[s] + H$_2$S         & SiS     \\
\hline   
11 & Fe[s]              & Fe 
       $\longrightarrow$ Fe[s]                  & Fe      \\ 
12 & solid iron         & FeO + H$_2$ 
       $\longrightarrow$ Fe[s] + H$_2$O         & FeO     \\
13 & (1)                & FeS + H$_2$ 
       $\longrightarrow$ Fe[s] + H$_2$S         & FeS     \\ 
14 &                    & Fe(OH)$_2$ + H$_2$ 
       $\longrightarrow$ Fe[s] + 2 H$_2$O       & Fe(OH)$_2$ \\ 
15 &                    & 2  FeH
       $\longrightarrow$ 2 Fe[s] + H$_2$         & FeH     \\ 
\hline 
16 & FeO[s]             & FeO 
       $\longrightarrow$ FeO[s]                  & FeO\\
17 & iron\,(II) oxide   & Fe + H$_2$O
       $\longrightarrow$ FeO[s] + H$_2$          & Fe\\
18 & (3)                & FeS + H$_2$O 
       $\longrightarrow$ FeO[s] + H$_2$S         & FeS\\
19 &                    & Fe(OH)$_2$
       $\longrightarrow$ FeO[s] + H$_2$          & Fe(OH)$_2$\\
20 &                    & 2 FeH + 2 H$_2$O
       $\longrightarrow$ 2 FeO[s] + 3 H$_2$      & FeH \\
\hline
21 & FeS[s]             & FeS
       $\longrightarrow$ FeS[s]                       & FeS\\
22 & iron sulphide      & Fe + H$_2$S
       $\longrightarrow$ FeS[s]     + H$_2$           & Fe\\
23 & (3)                & FeO + H$_2$S 
       $\longrightarrow$ FeS[s] + H$_2$O     & $\min\{$FeO, H$_2$S$\}$\\
24 &                    & Fe(OH)$_2$ + H$_2$S     
       $\longrightarrow$ FeS[s] + 2 H$_2$O   & $\min\{$Fe(OH)$_2$, H$_2$S$\}$\\
25 &                    & 2 FeH + 2 H$_2$S
       $\longrightarrow$ 2 FeS[s] + 3 H$_2$  & $\min\{$FeH, H$_2$S$\}$\\
\hline
26 & Fe$_2$O$_3$[s]     & 2 Fe + 3 H$_2$O 
       $\longrightarrow$ Fe$_2$O$_3$[s] + 3 H$_2$        & $\frac{1}{2}$Fe\\
27 & iron\,(III) oxide  & 2 FeO + H$_2$O
       $\longrightarrow$ Fe$_2$O$_3$[s] + H$_2$          & $\frac{1}{2}$FeO\\
28 & (3)                & 2 FeS + 3 H$_2$O
       $\longrightarrow$ Fe$_2$O$_3$[s] + 2 H$_2$S + H$_2$&$\frac{1}{2}$FeS\\
29 &                    & 2 Fe(OH)$_2$ 
       $\longrightarrow$ Fe$_2$O$_3$[s] + H$_2$O + H$_2$ & $\frac{1}{2}$Fe(OH)$_2$\\
30 &                    & 2 FeH + 3 H$_2$O
       $\longrightarrow$ Fe$_2$O$_3$[s] + 4 H$_2$ & $\frac{1}{2}$FeH\\
\hline
31 & MgO[s]             & Mg + H$_2$O 
      $\longrightarrow$ MgO[s] + H$_2$                & Mg\\
32 & periclase          & 2 MgH + 2 H$_2$O
      $\longrightarrow$ 2 MgO[s] + 3 H$_2$            & $\frac{1}{2}$MgH\\ 
33 & (3)                & 2 MgOH
      $\longrightarrow$ 2 MgO[s] + H$_2$              & $\frac{1}{2}$MgOH\\
34 &                    & Mg(OH)$_2$
      $\longrightarrow$ MgO[s] + H$_2$O               & Mg(OH)$_2$\\
\hline
35 & MgSiO$_3$[s]     & Mg + SiO + 2 H$_2$O 
     $\longrightarrow$ MgSiO$_3$[s] + H$_2$
                                  & $\min\{$Mg, SiO$\}$\\ 
36 & enstatite        & Mg + SiS + 3 H$_2$O 
     $\longrightarrow$ MgSiO$_3$[s] + H$_2$S + 2 H$_2$ 
                                  & $\min\{$Mg, SiS$\}$\\ 
37 & (3)              & 2 Mg + 2 SiH + 6 H$_2$O 
     $\longrightarrow$ 2 MgSiO$_3$[s] + 7 H$_2$
                                  & $\min\{$Mg, SiH$\}$\\ 
38 &                  & 2 MgOH + 2 SiO + 2 H$_2$O
     $\longrightarrow$ 2 MgSiO$_3$[s] + 3 H$_2$    
                                  & $\min\{\frac{1}{2}$MgOH, $\frac{1}{2}$SiO$\}$ \\
39 &                  & 2 MgOH + 2 SiS + 4 H$_2$O
     $\longrightarrow$ 2 MgSiO$_3$[s] + 2 H$_2$S + 3 H$_2$ 
                                  & $\min\{\frac{1}{2}$MgOH, $\frac{1}{2}$SiS$\}$ \\
40 &                  & MgOH + SiH + 2 H$_2$O
     $\longrightarrow$ MgSiO$_3$[s] + 3 H$_2$
                                  & $\min\{\frac{1}{2}$MgOH, $\frac{1}{2}$SiH$\}$ \\
41 &                  & Mg(OH)$_2$ + SiO 
     $\longrightarrow$ 2 MgSiO$_3$[s] +  H$_2$
                                  & $\min\{$Mg(OH)$_2$, SiO$\}$ \\ 
42 &                  & Mg(OH)$_2$ + SiS + H$_2$O
     $\longrightarrow$ MgSiO$_3$[s] + H$_2$S+ H$_2$
                                  & $\min\{$Mg(OH)$_2$, SiS$\}$ \\
43 &                  & 2 Mg(OH)$_2$ + 2 SiH + 2 H$_2$O
     $\longrightarrow$ 2 MgSiO$_3$[s] + 5 H$_2$
                                  & $\min\{$Mg(OH)$_2$, SiH$\}$ \\
44 &                  & 2 MgH +  2 SiO + 4 H$_2$O
     $\longrightarrow$ 2 MgSiO$_3$[s]+ 5 H$_2$
                                  & $\min\{$MgH, SiO$\}$ \\
45 &                  & 2 MgH +  2 SiS + 6 H$_2$O
     $\longrightarrow$ 2 MgSiO$_3$[s]+ 2 H$_2$S + 5 H$_2$
                                  & $\min\{$MgH, SiS$\}$ \\
46 &                  & MgH + SiH + 3 H$_2$O
     $\longrightarrow$  MgSiO$_3$[s]+ 4 H$_2$
                                  & $\min\{$MgH, SiH$\}$ \\
\hline
47 & Mg$_2$SiO$_4$[s] & 2 Mg + SiO + 3 H$_2$O
     $\longrightarrow$ Mg$_2$SiO$_4$[s] + 3 H$_2$  
                                  & $\min\{\frac{1}{2}$Mg, SiO$\}$\\
48 & forsterite       & 2 MgOH + SiO + H$_2$O
     $\longrightarrow$ Mg$_2$SiO$_4$[s] + 2 H$_2$
                                  & $\min\{\frac{1}{2}$MgOH, SiO$\}$\\ 
49 & (3)              & 2 Mg(OH)$_2$ + SiO 
     $\longrightarrow$ Mg$_2$SiO$_4$[s] + H$_2$O + H$_2$
                                  & $\min\{\frac{1}{2}$Mg(OH)$_2$, SiO$\}$ \\
50 &                  & 2 MgH + SiO + 3 H$_2$O
     $\longrightarrow$ Mg$_2$SiO$_4$[s] + 4 H$_2$
                                  & $\min\{\frac{1}{2}$MgH, SiO$\}$ \\
51 &                  & 2 Mg + SiS + 4 H$_2$O            
     $\longrightarrow$ Mg$_2$SiO$_4$[s] + H$_2$S + 3 H$_2$
                                  & $\min\{\frac{1}{2}$Mg, SiS\} \\
52 &                  & 2 MgOH + SiS + 2 H$_2$O 
     $\longrightarrow$ Mg$_2$SiO$_4$[s] + H$_2$S + 2 H$_2$
                                  & $\min\{\frac{1}{2}$MgOH, SiS\}\\
53 &                  & 2 Mg(OH)$_2$ + SiS 
     $\longrightarrow$ Mg$_2$SiO$_4$[s] + H$_2$ + H$_2$S
                                  & $\min\{\frac{1}{2}$Mg(OH)$_2$, SiS\} \\
54 &                  & 2 MgH + SiS + 4 H$_2$O
     $\longrightarrow$ Mg$_2$SiO$_4$[s] + H$_2$S + 4 H$_2$
                                  & $\min\{\frac{1}{2}$MgH, SiS$\}$ \\
55 &                  & 4 Mg + 2 SiH + 8 H$_2$O            
     $\longrightarrow$ 2 Mg$_2$SiO$_4$[s] + 9 H$_2$
                                  & $\min\{\frac{1}{2}$Mg, SiH\} \\
56 &                  & 4 MgOH + 2 SiH + 4 H$_2$O 
     $\longrightarrow$ 2 Mg$_2$SiO$_4$[s] + 7 H$_2$
                                  & $\min\{\frac{1}{2}$MgOH, SiH\}\\
57 &                  & 4 Mg(OH)$_2$ + 2 SiH 
     $\longrightarrow$ 2 Mg$_2$SiO$_4$[s] + 5 H$_2$
                                  & $\min\{\frac{1}{2}$Mg(OH)$_2$, SiH\} \\
58 &                  & 4 MgH + 2 SiH + 8 H$_2$O
     $\longrightarrow$ 2 Mg$_2$SiO$_4$[s] + 11 H$_2$
                                  & $\min\{\frac{1}{2}$MgH, SiS$\}$ \\
\hline 
59 & Al$_2$O$_3$[s]   & 2 Al + 3 H$_2$O 
     $\longrightarrow$ Al$_2$O$_3$[s] + 3 H$_2$   & $\frac{1}{2}$Al\\
60 & aluminia         & 2 AlOH + H$_2$O 
     $\longrightarrow$ Al$_2$O$_3$[s] + 2 H$_2$   & $\frac{1}{2}$AlOH \\ 
61 & (3)              &  2 AlH + 3 H$_2$O 
     $\longrightarrow$ Al$_2$O$_3$[s] + 4 H$_2$   & $\frac{1}{2}$AlH\\
62 &                  & Al$_2$O + 2 H$_2$O
     $\longrightarrow$ Al$_2$O$_3$[s] + 2 H$_2$   & Al$_2$O\\
63 &                  & 2 AlO$_2$H 
$\longrightarrow$ Al$_2$O$_3$[s] + H$_2$O    & $\frac{1}{2}$AlO$_2$H\\
\end{tabular}}
\tablefoot{The efficiency of the reaction is limited by the collision rate of the key species, which has the lowest abundance among the reactants. The notation $\frac{1}{2}$ in the r.h.s. column means that only every second collision (and sticking) event initiates one reaction. Data sources for the supersaturation ratios (and saturation vapor pressures): (1) \citet{Helling2006}; (2) \citet{Nuth2006}; (3) \citet{Sharp1990}.}
\end{table*}

\begin{table*}
\centering
\caption{Table~\ref{tab:chemreak} continued}
\label{tab:chemreak2}
\resizebox{14.0cm}{!}{
\begin{tabular}{c|c|l|l}
{\bf Index $r$} & {\bf Solid s} & {\bf Surface reaction} & {\bf Key species} \\
\hline
64 & CaTiO$_3$[s]         & Ca + Ti + 3 H$_2$O     
      $\longrightarrow$ CaTiO$_3$[s] + 3 H$_2$       & $\min\{$Ca, Ti$\}$\\  
65 & perovskite           & Ca + TiO + 2 H$_2$O 
      $\longrightarrow$ CaTiO$_3$[s] + 2 H$_2$       & $\min\{$Ca, TiO$\}$\\
66 & (3)                  & Ca + TiO$_2$ + H$_2$O 
      $\longrightarrow$ CaTiO$_3$[s] + H$_2$         & $\min\{$Ca, TiO$_2\}$\\
67 &                      & Ca + TiS + 3 H$_2$O 
      $\longrightarrow$ CaTiO$_3$[s] + H$_2$S + 2 H$_2$  & $\min\{$Ca, TiS$\}$\\
68 &                      & CaO + Ti + 2 H$_2$O 
      $\longrightarrow$ CaTiO$_3$[s] + 2 H$_2$       & $\min\{$CaO, Ti$\}$\\
69 &                      & CaO + TiO + H$_2$O 
      $\longrightarrow$ CaTiO$_3$[s] + H$_2$         & $\min\{$CaO, TiO$\}$\\
70 &                      & CaO + TiO$_2$
      $\longrightarrow$ CaTiO$_3$[s]                 & $\min\{$CaO, TiO$_2\}$\\
71 &                      & CaO + TiS + 2 H$_2$O 
      $\longrightarrow$ CaTiO$_3$[s] + H$_2$S + H$_2$ & $\min\{$CaO, TiO$\}$\\
72 &                      & CaS + Ti + 3 H$_2$O 
      $\longrightarrow$ CaTiO$_3$[s] + H$_2$S + H$_2$ & $\min\{$CaS, Ti$\}$\\
73 &                      & CaS + TiO + 2 H$_2$O 
      $\longrightarrow$ CaTiO$_3$[s] + H$_2$S + 2 H$_2$ &$\min\{$CaS, TiO$\}$\\
74 &                      & CaS + TiO$_2$ + H$_2$O 
      $\longrightarrow$ CaTiO$_3$[s] + H$_2$S        & $\min\{$CaS, TiO$_2\}$\\
75 &                      & CaS + TiS + 3 H$_2$O 
      $\longrightarrow$ CaTiO$_3$[s] + 2 H$_2$S + H$_2$ &$\min\{$CaS, TiO$\}$\\
76 &                      & Ca(OH)$_2$ + Ti + H$_2$O 
      $\longrightarrow$ CaTiO$_3$[s] + 2 H$_2$  & $\min\{$Ca(OH)$_2$, Ti$\}$\\
77 &                      & Ca(OH)$_2$ + TiO 
      $\longrightarrow$ CaTiO$_3$[s] + H$_2$    & $\min\{$Ca(OH)$_2$, TiO$\}$\\
78 &                      & Ca(OH)$_2$ + TiO$_2$ 
      $\longrightarrow$ CaTiO$_3$[s] + H$_2$O   &$\min\rm\{Ca(OH)_2,TiO_2\}$\\
79 &                      & Ca(OH)$_2$ + TiS + H$_2$O
      $\longrightarrow$ CaTiO$_3$[s] + H$_2$S + H$_2$   & $\min\{$Ca(OH)$_2$, TiO$\}$\\
80 &                      & 2 CaH + 2 Ti + 6 H$_2$O
      $\longrightarrow$ 2 CaTiO$_3$[s] + 7 H$_2$   &   $\min\{$CaH, Ti$\}$\\
81 &                      & 2 CaH + 2 TiO + 4 H$_2$O
      $\longrightarrow$ 2 CaTiO$_3$[s] + 5 H$_2$   &   $\min\{$CaH, TiO$\}$\\
82 &                      & 2 CaH + 2 TiO$_2$ + 2 H$_2$O
      $\longrightarrow$ 2 CaTiO$_3$[s] + 3 H$_2$   &   $\min\{$CaH, TiO$_2$ $\}$\\
83 &                      & 2 CaH + 2 TiS + 6 H$_2$O
      $\longrightarrow$ 2 CaTiO$_3$[s] + 2 H$_2$S +5 H$_2$   &   $\min\{$CaH, TiS$\}$\\
84 &                      & 2 CaOH + 2 Ti + 4 H$_2$O
      $\longrightarrow$ 2 CaTiO$_3$[s] + 5 H$_2$   &   $\min\{$CaOH, Ti$\}$\\
85 &                      & 2 CaOH + 2 TiO + 2 H$_2$O
      $\longrightarrow$ 2 CaTiO$_3$[s] + 3 H$_2$   &   $\min\{$CaOH, TiO$\}$\\
86 &                      & 2 CaOH + 2 TiO$_2$
      $\longrightarrow$ 2 CaTiO$_3$[s] + H$_2$   &   $\min\{$CaOH, TiO$_2$ $\}$\\
87 &                      & 2 CaOH + 2 TiS + 4 H$_2$O
      $\longrightarrow$ 2 CaTiO$_3$[s] + 2 H$_2$S + 3 H$_2$   &   $\min\{$CaOH, TiS$\}$\\
\hline
\end{tabular}}
\end{table*}
\clearpage
\section{Dust optical constants data references}
\label{ap:dust}
\begin{table}[hbt!]
\renewcommand{\arraystretch}{1.2}
\caption{Dust optical data sources. For details on the wavelength ranges please refer to Table 1 in \citet{Kitzmann2018}.}             % title of Table
\label{tab:dust}      % is used to refer this table in the text
\centering
                  % used for centering table
\begin{tabular}{l l} % centered columns (4 columns)
\hline\hline          % inserts double horizontal lines
\noalign{\smallskip}
Condensate & Reference \\    % table heading
\hline                        % inserts single horizontal line
\noalign{\smallskip}
\ce{Al2O3} [s]&  \citet{AlO3_Begemann}$^{*t}$; \citet{Al2O3_Koike}$^{t}$ \\
CaTiO$_3$[s] & \citet{CaTiO3_Posch}$^{t}$; \citet{CaTiO3_Ueda}$^{f}$\\
Fe[s]&  Lynch \& Hunter in \citet{Palik1991}$^{t}$ \\ 
Fe$_2$O$_3$[s]&  A.H.M.J. Triaud$^{*t}$ \\ 
FeO[s]&  \cite{FeO_Henning}$^{*t}$ \\ 
FeS[s]&  \cite{FeS_Pollack}$^{t}$; \cite{FeS_Henning}$^{t}$ \\ 
Mg$_2$SiO$_4$[s]&  \cite{Mg2SiO4_Jager}$^{*t}$ \\ 
MgSiO$_3$[s]&  \cite{Mg2SiO4_Jager}$^{*t}$ \\ 
MgO[s]& Roessler \& Huffman in \citet{Palik1991}$^{t}$ \\
SiO[s] & Philipp in \citet{Palik1985}$^{t}$ \\
SiO$_2~$[s] & \citet{FeS_Henning}$^{*t}$; Philipp in \citet{Palik1985}$^{t}$\\
TiO$_2~$[s] & \citet{TiO2_Zeidler}$^{*t}$; \citet{CaTiO3_Posch}$^{*t}$ \\
\noalign{\smallskip}
\hline                                   %inserts single line
\end{tabular}
\tablefoot{We note we use the amorphous (sol-gel) data for Mg$_2$SiO$_4$[s] and MgSiO$_3$[s], and the amorphous data for SiO$_2$[s].\\
*Data from the Database of Optical Constants for Cosmic Dust, Laboratory Astrophysics Group of the AIU Jena.\\
$^t$Data from a printed or digital table.\\
$^f$Data from a figure.}
\end{table}
\clearpage
\section{Models with SiO nucleation}
\label{ap:SiO}
\begin{figure*}[hbt!]
   \centering
   \includegraphics[width=0.5\linewidth]{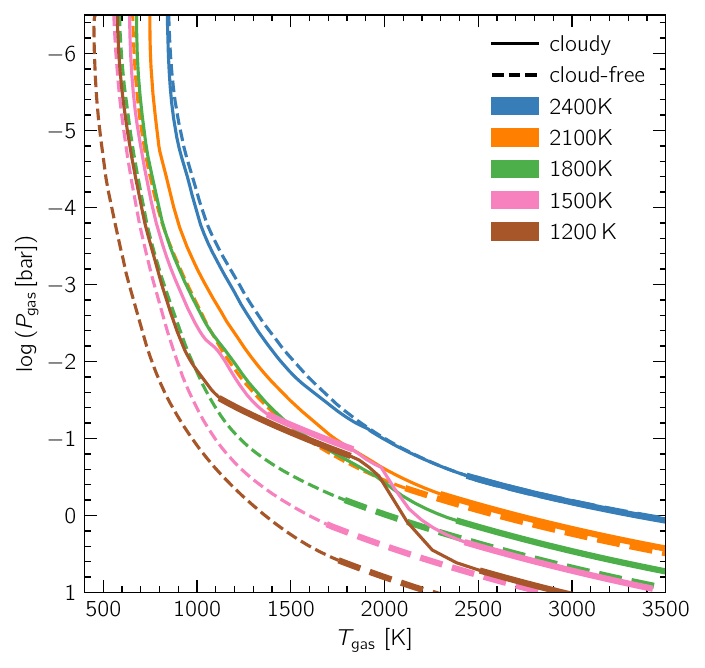}
    \caption{Pressure-temperature profiles for cloud-free \texttt{MSG} models (dashed curves) and cloudy \texttt{MSG} models with SiO nucleation (solid curves), at different effective temperatures and $\log(g)=4.0$. Convective zones are plotted with thicker lines while radiative zones are plotted with thinner lines.}
    \label{fig:pt_profiles_SiO_all}
\end{figure*}
\begin{figure*}[hbt!]
\begin{minipage}[b]{0.33\linewidth}
     \includegraphics[width=\linewidth, valign=t]{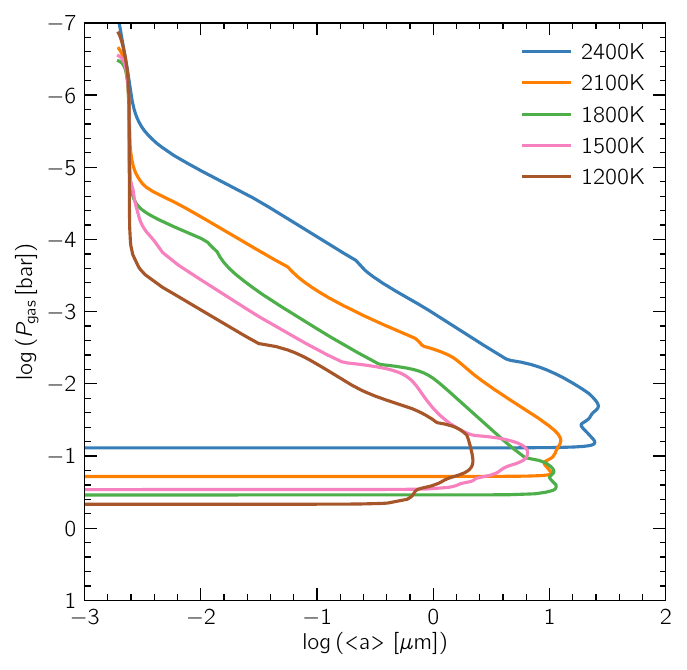}
\end{minipage}
\begin{minipage}[b]{0.33\linewidth}
     \includegraphics[width=\linewidth, valign=t]{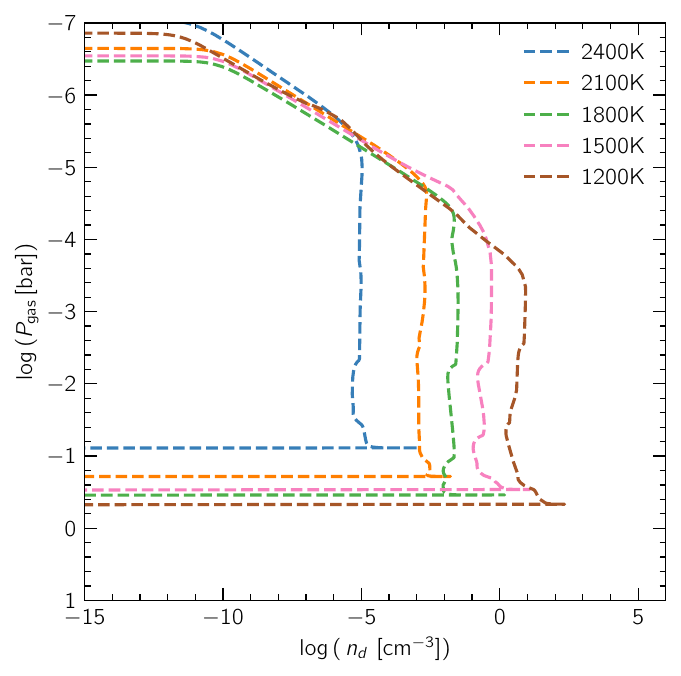}
\end{minipage}
\begin{minipage}[b]{0.33\linewidth}
     \includegraphics[width=\linewidth, valign=t]{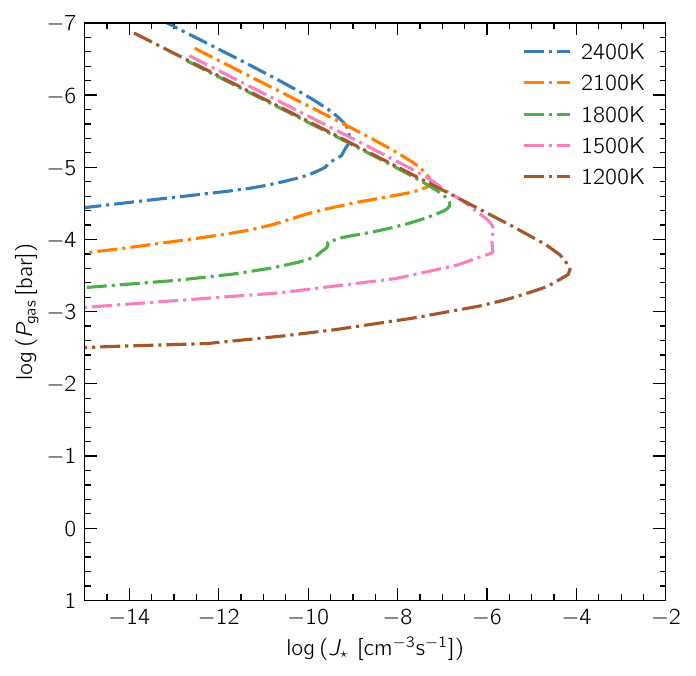}
\end{minipage}
     \caption{The average cloud particle size $\langle a \rangle$ (left), the cloud particle number density $n_d$ (middle) and the nucleation rate $J_\star$ (right) along the atmosphere for models with SiO nucleation at $T_{\rm{eff}}$ $=$ 2400\,K, 2100\,K, 1800\,K, 1500\,K and 1200\,K and $\log(g)=4.0$. The corresponding $P_{\mathrm{gas}}-T_{\mathrm{gas}}$ profiles are shown in Fig.~\ref{fig:pt_profiles_SiO_all}.}
     \label{fig:cloud_size_nd_SiO}
\end{figure*}
\clearpage
\section{The effect of decreasing the mixing efficiency - models with \ce{SiO} nucleation}
\label{ap:SiO_mix}
\begin{figure*}[hbt!]
\begin{minipage}[b]{0.5\linewidth}
   \includegraphics[width=\linewidth, valign=t]{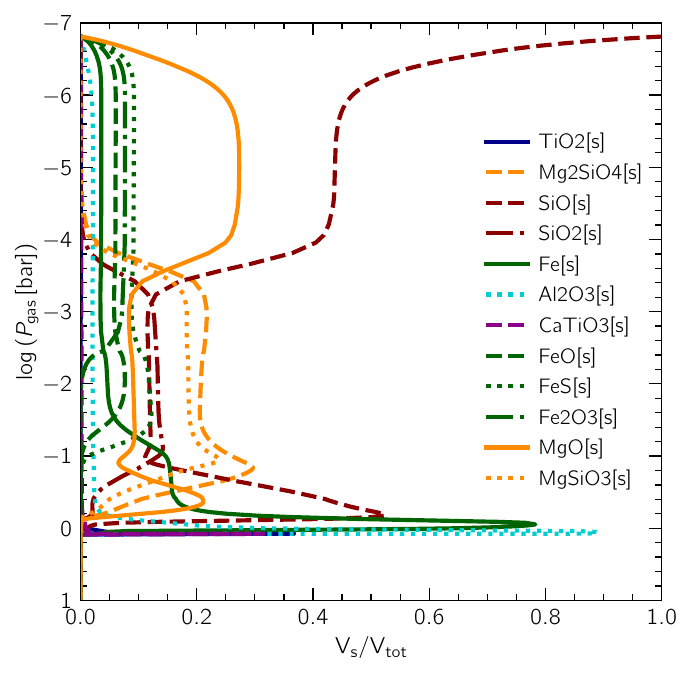}
\end{minipage}
\begin{minipage}[b]{0.5\linewidth}
   \includegraphics[width=\linewidth, valign=t]{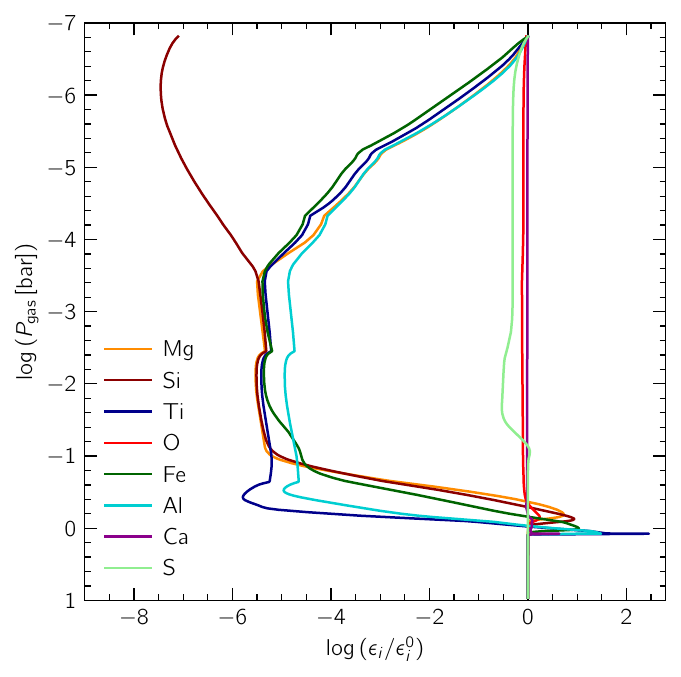}
\end{minipage}
     \caption{Composition of the cloud particles in units of volume fractions $V_s/V_{\mathrm{tot}}$ (left) and the relative gas-phase element depletions $\epsilon_i/ \epsilon_i^0$ (right) for a \texttt{MSG} model with SiO nucleation, at $T_{\rm{eff}}$$\,=\,$1500\,K and $\log(g)=4.0$ and with a 1000 times slower mixing timescale.}
     \label{fig:cloud_comp_SiO_mix}
\end{figure*}

\clearpage
\section{The effect of decreasing the mixing efficiency - models with \ce{TiO2} nucleation}
\label{ap:TiO2_mix}
\begin{figure}[hbt!]
    \centering
    \includegraphics[width=0.5\linewidth]{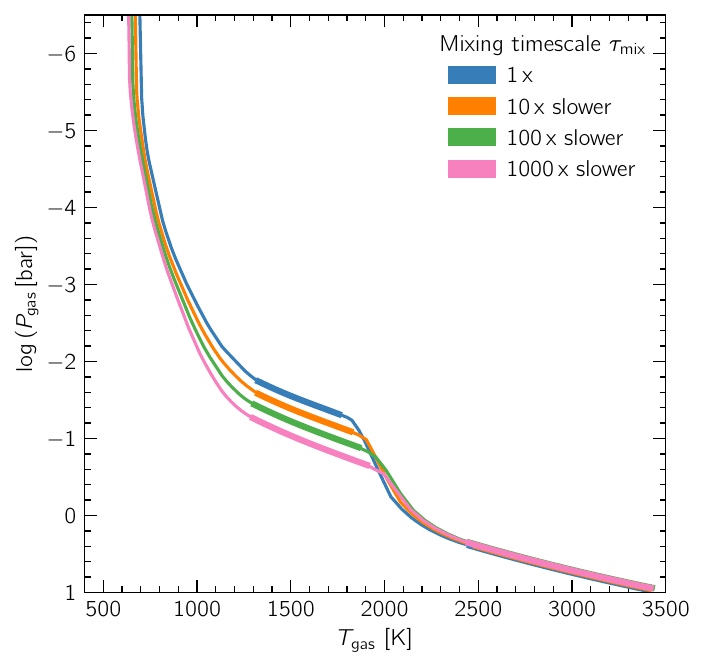}
    \caption{Pressure-temperature profiles for \texttt{MSG} models with TiO$_2$ nucleation, at $T_{\rm{eff}}$$\,=\,$1500\,K and $\log(g)=4.0$, and different mixing timescale scalings (1x, 10x slower, 100x slower, 1000x slower). Convective zones are plotted with thicker lines while radiative zones are plotted with thinner lines.}
    \label{fig:TiO2_mix_pt}
\end{figure}
\begin{figure*}[hbt!]
\begin{minipage}[b]{0.33\linewidth}
    \includegraphics[width=\linewidth, valign=t]{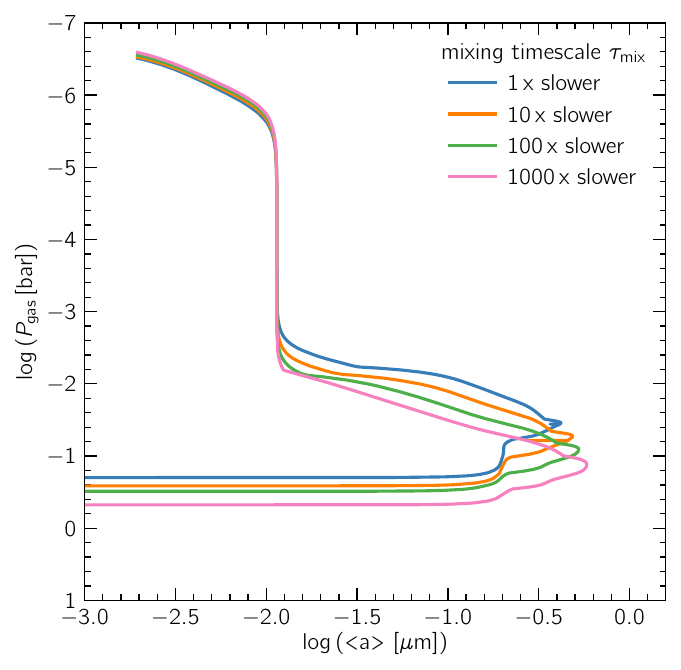}
\end{minipage}
\begin{minipage}[b]{0.33\linewidth}
    \includegraphics[width=\linewidth, valign=t]{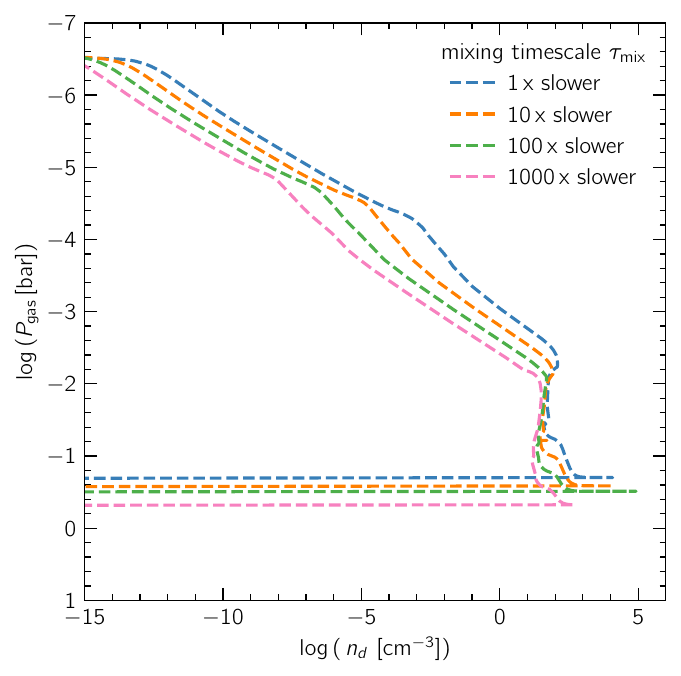}
\end{minipage}
\begin{minipage}[b]{0.33\linewidth}
    \includegraphics[width=\linewidth, valign=t]{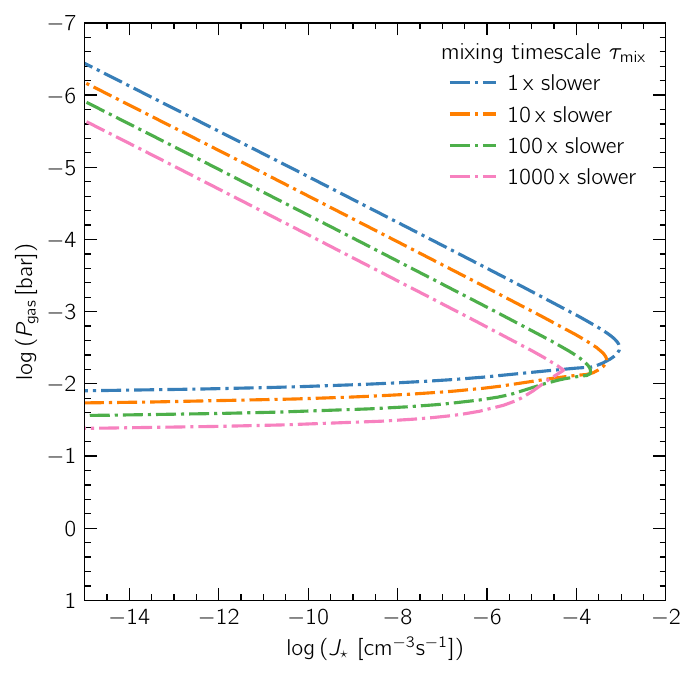}
\end{minipage}
\caption{The average cloud particle size $\langle a \rangle$ (left), the cloud particle number density $n_d$ (middle) and the nucleation rate $J_\star$ (right) along the atmosphere for models with TiO$_2$ nucleation at $T_{\rm{eff}}$ $=$ 1500\,K and $\log(g)=4.0$, and different mixing timescale scalings (1x, 10x slower, 100x slower, 1000x slower).}
\label{fig:TiO2_mix}
\end{figure*}
\begin{figure*}
\begin{minipage}[b]{0.5\linewidth}
   \includegraphics[width=\linewidth, valign=t]{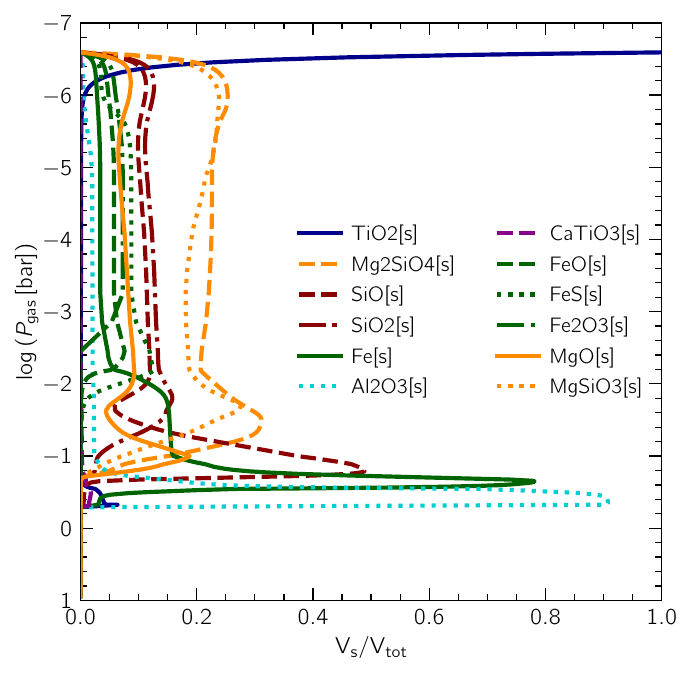}
\end{minipage}
\begin{minipage}[b]{0.5\linewidth}
   \includegraphics[width=\linewidth, valign=t]{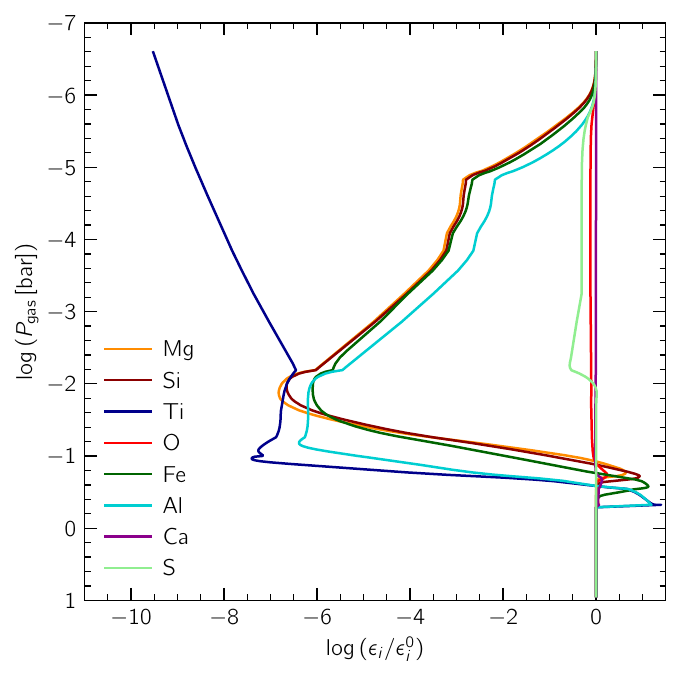}
\end{minipage}
     \caption{Composition of the cloud particles in units of volume fractions $V_s/V_{\mathrm{tot}}$ (left) and the relative gas-phase element depletions $\epsilon_i/ \epsilon_i^0$ (right) for a \texttt{MSG} model with TiO$_2$ nucleation, at $T_{\rm{eff}}$$\,=\,$1500\,K and $\log(g)=4.0$ and with a 1000 times slower mixing timescale.\\
     \\}
     \label{fig:cloud_comp_TiO2_mix}
\end{figure*}

\begin{figure}[hbt!]
\centering
\includegraphics[width=0.48\linewidth]{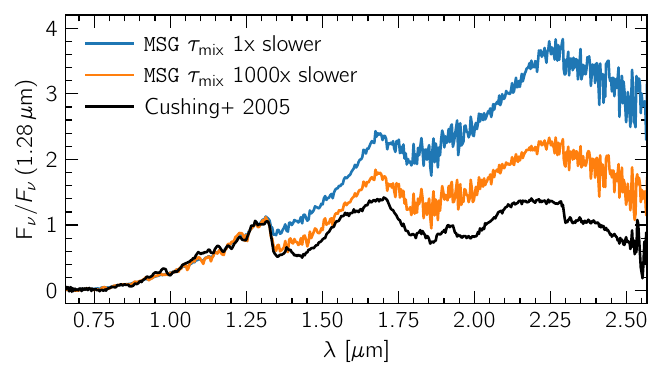}
\includegraphics[width=0.48\linewidth]{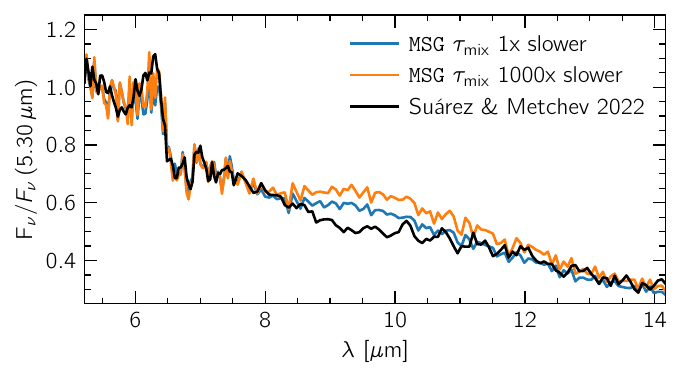}
\caption{Comparison between observed spectra of 2MASS 1507-1627 (black curves) \citep{Cushing2005, Suarez2022} and two \texttt{MSG} cloudy models with \ce{TiO2} nucleation: one with fully self-consistent mixing (blue curves), and the other where the mixing timescale was scaled up by 1000 times (orange curves). The \texttt{MSG} models are at \Teff=1600\,K and log(g)=4.0.}
\label{fig:spectra_TiO2_mix}
\end{figure}
\end{appendix}
\end{document}